\documentclass[aip,apl,reprint,noshowpacs,noshowkeys,amssymb,amsmath,amsfonts,bibnotes]{revtex4-2}

\makeatletter
\let\@fnsymbol\@fnsymbol@latex
\@booleanfalse\altaffilletter@sw
\def\frontmatter@makefnmark{\@textsuperscript{\normalfont\@thefnmark}}
\makeatother

\usepackage{graphicx}
\usepackage{theorem}
\usepackage{dcolumn}
\usepackage{longtable}
\usepackage{bm}
\usepackage{tabularx}
\usepackage{multirow}
\usepackage{color}
\usepackage[normalem]{ulem}
\usepackage{enumitem}

\usepackage{siunitx}

\newcommand{\SSS}{\scriptscriptstyle}
\newcommand{\DS}{\displaystyle}

\newcommand{\ie}{{\textit{i.e.}}}
\newcommand{\eg}{{\textit{e.g.}}}
\newcommand{\etal}{{\textit{et al.}}}
\newcommand{\etc}{{\textit{etc.}}}

\newcommand{\Dd}{{\rm d}}

\newcommand{\muB}{{\mu_{\SSS\text{B}}}}

\newcommand{\aB}{a_{\SSS\text{B}}}
\newcommand{\KB}{k_{\SSS\text{B}}}
\newcommand{\EF}{E_{\SSS\text{F}}}

\def\imgpath{./}

\renewcommand{~}{\,}

\definecolor{midnightblue}{rgb}{0.1, 0.1, 0.44}
\definecolor{oucrimsonred}{rgb}{0.6, 0.0, 0.0}

\renewcommand{\thetable}{\arabic{table}}

\makeatletter
\renewcommand{\figurename}{Fig.}
\renewcommand{\tablename}{Table}
\renewcommand{\fnum@figure}[1]{\textbf{\figurename~\thefigure:~}}
\renewcommand{\fnum@table}[1]{\textbf{\tablename~\thetable:~}}
\makeatother

\usepackage{makecell}

\begin{document}

\title{Quantum Electronics on Quantum Liquids and Solids}

\author{Wei Guo}\email{wguo@eng.famu.fsu.edu}
\affiliation{National High Magnetic Field Laboratory, 1800 East Paul Dirac Drive, Tallahassee, Florida 32310, USA\looseness=-1}
\affiliation{Department of Mechanical Engineering, FAMU-FSU College of Engineering, Florida State University, Tallahassee, Florida 32310, USA\looseness=-1}

\author{Denis Konstantinov}\email{denis@oist.jp}
\affiliation{Quantum Dynamics Unit, Okinawa Institute of Science and Technology, Onna, 904-0412, Okinawa, Japan\looseness=-1}

\author{Dafei Jin}\email{dfjin@nd.edu}
\affiliation{Department of Physics and Astronomy, University of Notre Dame, Notre Dame, Indiana 46556, USA\looseness=-1}

\date{\today}

\begin{abstract}

Nonpolar atoms or molecules with low particle mass and weak inter-particle interactions can form quantum liquids and solids (QLS) at low temperatures. Excess electrons naturally bind to the surfaces of QLS in a vacuum, exhibiting unique quantum electronic behaviors in two and lower dimensions. This article reviews the historical development and recent progress in this field. Key topics include collective and individual electron transport on liquid helium, solid neon, and solid hydrogen; theoretical proposals and experimental efforts toward single-electron qubits on superfluid helium; the recent experimental realization of single-electron charge qubits on solid neon; and related theoretical calculations. Finally, we discuss and envision future exploration of quantum electronics in heterogeneous QLS systems.

\end{abstract}

\maketitle
\pretolerance=9000 

\section{Overview}

\subsection{Notion of quantum liquids and solids}

\begin{figure}[htb]
	\includegraphics[scale=0.8]{\imgpath/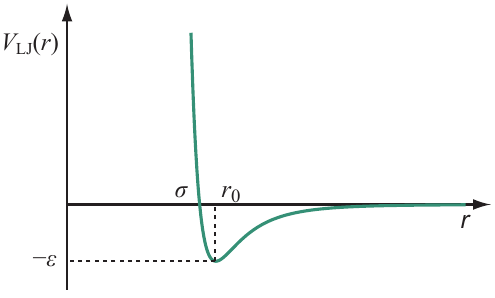}
	\caption{\label{Fig:LJPotential} Lennard-Jones (LJ) potential between nonpolar atoms and molecules.}
\end{figure}

Quantum liquids and solids (QLS) are substances whose behaviors show appreciable deviation from those of classical liquids and solids due to the quantum nature of constituent particles. QLS typically comprise nonpolar particles (atoms or molecules) with a low particle mass and weak inter-particle interaction. The interaction can be well described by a Lennard-Jones (LJ) potential,~\cite{pollack1964solid,Klein1976RareGas,silvera1980solid,cazorla2017simulation}
\begin{equation}
	V(r) = 4\varepsilon \left[ \left( \frac{\sigma}{r} \right)^{12} - \left( \frac{\sigma}{r} \right)^{6}  \right],
\end{equation}
where $r$ is the variable inter-particle distance, $\sigma$ and $\varepsilon$ are, respectively, the characteristic length and energy, obtained from curve fitting for a given particle species. The LJ potential is short-range repulsive and long-range attractive. With increasing $r$ from zero, $V(r)$ changes its sign from positive (repulsive) to negative (attractive) at $r=\sigma$, \ie, $V(\sigma)=0$, and reaches its minimum $-\varepsilon$ at $r=r_0 \equiv 2^{1/6}\sigma \approx 1.122\sigma$, \ie, $V(r_0)=-\varepsilon$. (See Fig.~\ref{Fig:LJPotential}.)

The \emph{quantumness} of a substance can be quantified by the de Boer parameter $\mathit{\Lambda}$, which was first introduced by de Boer and co-worker in the context of the \emph{Quantum Theorem of Corresponding States}.~\cite{de1948quantum1,de1948quantum2} It is defined as the ratio between the de Broglie wavelength $\lambda$ of the relative motion of two particles and the mean distance $d$ between the two particles.~\cite{pollack1964solid,Klein1976RareGas,silvera1980solid,cazorla2017simulation} For nonpolar particles interacting through an LJ potential, 
\begin{equation}
	\mathit{\Lambda} \equiv \frac{\lambda}{d} = \frac{h/\sqrt{m\varepsilon_{\text{k}}}}{d} \approx \frac{h}{\sigma\sqrt{m\varepsilon}}, \label{Eqn:deBoerParameter}
\end{equation}
where $h$ is the Planck constant, $m$ is the particle mass, $d\approx \sigma$ is the approximated mean distance, and $\varepsilon_{\text{k}}\approx \varepsilon$ is the approximated zero-point kinetic energy of the relative motion of two particles.

For each particle species, $\mathit{\Lambda}$ can be calculated with the known $m$ and commonly adopted $\sigma$ and $\varepsilon$. Table~\ref{Table1} shows the calculated $\mathit{\Lambda}$ of representative nonpolar particles, including atomic helium-3 ($^3$He), helium-4 ($^4$He), neon (Ne), and molecular hydrogen (H$_2$), hydrogen deuteride (HD), deuterium (D$_2$). Noble-element $^3$He has the largest $\mathit{\Lambda}$, followed by noble-element $^4$He, non-noble-element H$_2$, HD, and D$_2$, and then noble-element Ne. Natural Ne consists of three stable isotopes $^{20}$Ne, $^{21}$Ne, and $^{22}$Ne, which have only a small fraction of mass difference, resulting in nearly the same  $\mathit{\Lambda}$ for all of them. 

\begin{table}[hbt]
	\caption{Particle mass $m$, characteristic length $\sigma$, energy $\varepsilon$ in the Lennard-Jones potentials, and calculated de Boer parameter $\mathit{\Lambda}$ of representative nonpolar particles.~\cite{mason1954intermolecular,knaap1961lennard,hansen1968ground,ebner1979density} \label{Table1}}
	\centering
	\begin{ruledtabular}
		\begin{tabular}{ccccc}			
			Particle		& $m$ (amu) & $\sigma$ (\AA)		& $\varepsilon$ (K)     & $\mathit{\Lambda}$         \\
			\hline
			Atomic $^3$He    		& 3.0160 & 2.556				& 10.2                & 3.09               \\
			Atomic $^4$He         	& 4.0026 & 2.556   			& 10.2                & 2.68               \\
			Atomic Ne              & 20.180 & 2.749              & 35.6                 & 0.59               \\
			\hline
			Molecular H$_2$      	 	& 2.0157 & 2.928				& 37.0				   & 1.73               \\
			Molecular HD      	 	& 3.0219 & 2.928				& 37.0				   & 1.41               \\
			Molecular D$_2$           & 4.0282 & 2.928              & 37.0                 & 1.22               
		\end{tabular}
	\end{ruledtabular}
\end{table}

A large de Boer parameter $\mathit{\Lambda}$ indicates significant quantum fluctuations at low temperatures. For instance, $^3$He and $^4$He remain to be liquid even as temperature approaches absolute zero unless an external pressure of 34~bar for $^3$He and 25~bar for $^4$He is applied to drive the liquid-solid transition.~\cite{Wilks1987} Also, liquid $^4$He undergoes a second-order phase transition from the normal liquid phase into the superfluid phase at about 2.17~K under saturated vapor pressure. Liquid $^3$He undergoes a superfluid phase transition at a much lower temperature around 2.5~mK when fermionic $^3$He atoms form Cooper pairs.~\cite{volovik2003universe} Parahydrogen ($p$H$_2$) with antiparallel proton spins is theoretically predicted to exhibit superfluidity at low temperatures but practically hindered by solidification.~\cite{silvera1980solid,sindzingre1991superfluidity,apenko1999critical,osychenko2012superfluidity,boninsegni2018search}

Quantum solids helium, hydrogen, and neon exhibit distinct behaviors from those of classical solids. For example, the spatial fluctuations of solid $^4$He atoms can reach up to 30\% of the interatomic distances due to their substantial zero-point energies.~\cite{cazorla2017simulation} The high mobility of atoms and vacancies in quantum solids contribute to the so-called ``anomalous" plasticity.~\cite{kim2004observation,day2007low} This allows these materials to deform under stress in unusual manners. Such plasticity has been observed in $^4$He and has implications for understanding mechanical behaviors at low temperatures.~\cite{kim2004observation,day2007low} Moreover, the anharmonic potential in these systems results in deviations from simple harmonic behavior, influencing thermal and acoustic properties significantly.~\cite{ceperley1995path}

\subsection{Surface electronic states}

The nonpolar particles listed in Table~\ref{Table1} share some common properties. They have tightly bound core orbitals filled with electrons in the ground state. The energy to excite an electron from the core to an outer orbital is notably high. These features give rise to unique electronic surface states in the condensed liquid and solid phases of these particles.~\cite{cole1969image,cole1970properties,shikin1970motion}

Consider an incoming excess electron being scattered off by an isolated nonpolar particle in a vacuum. The scattering is governed by a competition between the short-range repulsion and long-range attraction.~\cite{mott1933theory} The repulsion stems from the Pauli exclusion, which requires that the excess electron's wavefunction remains orthogonal to the filled core orbitals of the particle. This requirement causes the excess electron's wavefunction to oscillate rapidly within the core region of the particle, adding a significant positive contribution to the energy.~\cite{cohen1961cancellation} Meanwhile, at a greater distance, the excess electron's electric field weakly polarizes the particle, inducing an electric dipole moment that attracts the excess electron.~\cite{kivel1959elastic} Theoretical and experimental studies on the low-energy electron scattering cross-section, $\sigma_s=4\pi a^2_s$, have revealed that the scattering length $a_s$ is positive for the listed particles in Tables~\ref{Table1} and \ref{Table2},~\cite{mott1933theory} representing a net short-range repulsion. In contrast, for other closed-shell particles such as argon (Ar), krypton (Kr), and xenon (Xe) atoms, their $a_s$ is negative,~\cite{mott1933theory,golden1966comparison,ebner1979density,subramanian1987total} due to their larger atomic polarizability that produces a net attraction to the excess electron.

Extending this picture to an excess electron interacting with a continuum of a substance. The short-range repulsion between the substance particles and the excess electron creates an effective potential barrier $V_0$, preventing the excess electron from penetrating the substance. In the weak-scattering approximation ($n^{1/3}a_s\ll1$), the barrier height can be estimated as 
\begin{equation}
	V_0 \approx \frac{2\pi \hbar^2 na_s}{m_e},
\end{equation}
where $n$ is the particle number density of the substance and $m_e$ is the electron mass.~\cite{foldy1945multiple,levine1967mobility} For a liquid or solid phase with a high $n$, the Wigner-Seitz model,~\cite{Kittel1963-book} which accounts for multi-scattering processes,~\cite{lax1951multiple} yields a more accurate $V_0$. Away from the substance surface, the excess electron experiences an attractive polarization force from all the particles in the substance. This leads to an image potential,~\cite{Landau-book}
\begin{equation}
	V_{\text{p}}(z)=-\frac{(\epsilon-1)}{(\epsilon+1)}\frac{e^2}{4 z}, \label{Eqn:PolPotential}
\end{equation}
where $\epsilon$ is the relative dielectric constant of the substance, and $z$ is the vertical distance from a flat surface. Upon summarizing prior theoretical and experimental works,~\cite{bekefi1958collision,lee1961dielectric,springett1968stability,cole1970properties,constable1975dielectric,kierstead1976dielectric,grimes1976spectroscopy,ebner1979density,johnson1978electron,silvera1980solid,martini1991electron,cheng1994binding,collin2002microwave,jin2012finite} we list the recommended values of particle number density $n$, scattering length $a_s$, dielectric constant $\epsilon$, and barrier height $V_0$ in Table~\ref{Table2}. It is worthwhile to note that $V_0$ is on the order of 1~eV for all the listed substances. 

\begin{figure}[htb]
	\includegraphics[scale=0.5]{\imgpath/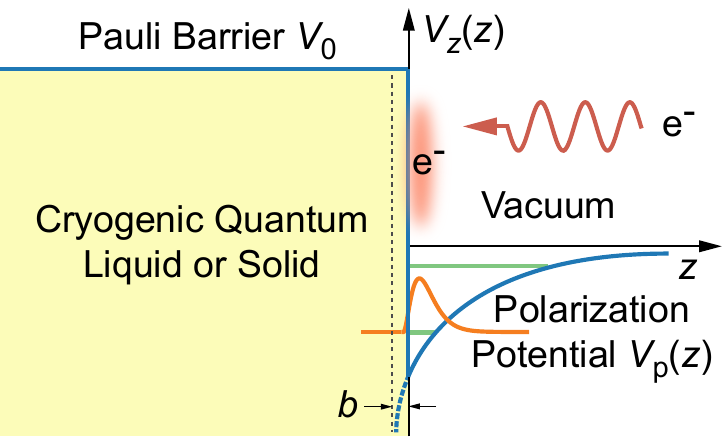}
	\caption{\label{Fig:SurfaceElectron} Schematic of the Pauli barrier, polarization potential, and surface states of a single electron on a cryogenic quantum liquid or solid.}
\end{figure}

The polarization potential Eq.~(\ref{Eqn:PolPotential}) diverges as $z\rightarrow0$. This divergence makes the calculated electronic states and energies sensitive to the barrier height and cutoff distance,~\cite{cole1970properties} unless the barrier is approximated with an infinite height and the electron wavefunction is strictly forced to zero at $z=0$. In practice, a ``regularized" potential with no divergence can be used,~\cite{hipolito1978electron,stern1978image}
\begin{equation}
	V_z(z) = \begin{cases}
		V_0, &\quad z<0, \\
	\DS	-\frac{(\epsilon-1)}{(\epsilon+1)}\frac{e^2}{4 (z+b)}, & \quad z>0,
	\end{cases}
\end{equation}
where the $z=0$ surface is redefined at a small distance $b$ away from the core of the top-layer particles. This is physically reasonable, because the near-core polarization attraction has already been counted in the scattering length $a_s$. So the simplest choice of $b$ is $a_s$. The dual effects of the repulsive barrier $V_0$ for $z<0$ and the attractive image potential $V_\text{p}(z)$ for $z>0$ can bind an excess electron to the surface of the substance. Figure~\ref{Fig:SurfaceElectron} shows schematically the $V_z
(z)$ profile and a trapped electron near a QLS surface.

\begin{table*}[htb]
	\caption{Particle number density $n$, s-wave scattering length $a_s$, dielectric constant $\epsilon$, potential barrier $V_0$, and calculated results of eigenenergies $E_{z,1}$, $E_{z,2}$, transition energy $\Delta E_{1\rightarrow 2}$, transition frequency $f_{1\rightarrow2}$, the mean electron-to-surface distances $\langle z\rangle_1$ and $\langle z\rangle_2$, for the surface electronic states on representative quantum liquids and solids at zero temperature and under zero pressure.~\cite{bekefi1958collision,lee1961dielectric,springett1968stability,cole1970properties,constable1975dielectric,kierstead1976dielectric,grimes1976spectroscopy,ebner1979density,johnson1978electron,silvera1980solid,crompton1982comment,zav1988quantum,martini1991electron,cheng1994binding,collin2002microwave,jin2012finite} \label{Table2}}
	\centering
	\begin{ruledtabular}
		\begin{tabular}{c|cccccccccc}
			Substance & $n$~({\AA}$^{-3}$) & $a_s$~(\AA) & $\epsilon$ & $V_0$~(eV) & $E_{z,1}$~(meV) & $E_{z,2}$~(meV) & $\Delta E_{1\rightarrow 2}$~(K) & $f_{1\rightarrow 2}$~(THz) & $\langle z \rangle_1 $~(nm) & $\langle z \rangle_2 $~(nm) \\
			\hline
			Liquid $^3$He & 0.0164 & 0.62 & 1.042 & 0.9 & $-0.382$ & $-0.093$ & 3.4 & 0.070 & 14.5 & 59.9 \\
			Liquid $^4$He & 0.0218 & 0.62 & 1.056 & 1.1 & $-0.676$ & $-0.163$ & 5.9 & 0.124 & 10.8 & 45.0 \\
			Solid Ne & 0.0460 & 0.38 & 1.244 & 0.7 & $-17.4$ & $-3.24$ & 165 & 3.43 & 1.66 & 9.04 \\
			\hline
			Solid H$_2$ & 0.0266 & 0.66 & 1.290 & 1.7 & $-16.5$ & $-3.74$ & 148 & 3.08 & 2.01 & 9.09 \\
			Solid HD & 0.0293 & 0.66 & 1.302 & 1.9 & $-17.4$ & $-3.98$ & 156 & 3.24 & 1.97 & 8.84 \\
			Solid D$_2$ & 0.0308 & 0.66 & 1.341 & 2.1 & $-21.3$ & $-4.89$ & 191 & 3.97 & 1.78 & 7.97 \\
		\end{tabular}	
	\end{ruledtabular}
\end{table*}

The eigenenergies $E_{z,n}$ and eigenfunctions $\psi_{z,n}(z)$ of an electron perpendicular to the surface satisfy the one-dimensional (1D) Schr\"{o}dinger equation,
\begin{equation}
	-\frac{\hbar^2}{2m_e}\frac{\partial^2}{\partial z^2}\psi_{z, n}(z)+V_z(z)\psi_{z, n}(z) = E_{z, n}\psi_{z, n}(z).
	\label{Wei-Eq1}
\end{equation}
This equation resembles the s-wave radial equation for an electron in the Coulomb field of charge $Ze=(\epsilon-1)e/4(\epsilon+1)$. To gain a qualitative insight, $V_0$ can be approximated as infinitely large. $E_{z, n}$ then takes an analytical form with a hydrogenic spectrum,~\cite{Kittel1963-book} 
\begin{equation}
	E_{z,n}=-\frac{Z^2e^4m_e}{2\hbar^2n^2},\quad n = 1, 2, 3, \dots
\end{equation}
In reality, the finite barrier height $V_0$ allows the electron's wavefunction to slightly leak into the substance at $z<0$, resulting in modifications to $E_{z, n}$.~\cite{cole1970properties} The more accurate $E_{z,1}$, $E_{z,2}$, the transition energy $\Delta E_{1\rightarrow 2}$, the transition frequency $f_{1\rightarrow 2}$, the mean electron-to-surface distance $\langle z\rangle_1= \langle \psi_{z,1}| z|\psi_{z,1}\rangle$ and $\langle z\rangle_2= \langle \psi_{z,2}| z|\psi_{z,2}\rangle$ are listed in Table~\ref{Table2}. As can be seen, the binding energy defined as $|E_{z,1}|$ ranges from $\sim$0.4~meV to $\sim$20~meV, corresponding to activation temperatures from $\sim$4.6~K to $\sim$232~K. Most experiments have been performed around 1~K or below.~\cite{kawakami2019image,bradbury2011efficient,yang2016coupling,koolstra2019coupling,zhou2022single,zhou2024electron} At such temperatures, the electron’s motion perpendicular to the surface is practically frozen in the $n=1$ ground state.

\subsection{Lateral motion of the surface electrons}

While an excess electron can be confined perpendicular to the surface of a QLS, it can also move along the surface. Due to the absence of usual impurities in classical materials that can cause disruptive scattering, the mobility of surface electrons on quantum liquids and solids can be exceptionally high. For instance, Sommer and Tanner reported an electron mobility of the order of 10$^6$~cm$^2$/Vs on superfluid $^4$He at $\sim$1~K,~\cite{sommerPRL1971tanner} limited by collisions of the electrons with helium atoms in the vapor. At lower temperatures, electron mobility as high as $10^8$~cm$^2$/Vs on liquid $^3$He and $^4$He was reported,~\cite{shirahamaJLTP1996mobility} limited by collisions with liquid surface excitations called ripplons.~\cite{shikin1974interaction,gaspari1975electron} The high mobility of electrons on a superfluid helium surface has been recognized as a promising feature for quantum electronic devices that require swift electron transport.~\cite{leiderer1992electrons,platzman1999quantum,bradbury2011efficient,papageorgiou2005counting,kawakami2023blueprint}

Mobility of electrons bound to solid hydrogen and solid neon surfaces has also been measured extensively, revealing intricate details about their behaviors and scattering mechanisms. On solid hydrogen, Troyanovskii and Khaikin found that electron mobility is primarily determined by scattering from surface defects at temperatures below 10~K, with minimal contributions from gas molecules or Rayleigh waves.~\cite{troyanovskii1981electron} Their measurements showed that the mobility follows a temperature dependence of $\mu\propto T^{-1}$, suggesting that the dominant scattering mechanism is from microscopic surface defects with dimensions on the order of the crystal cell size, around $5\times10^{-8}$~cm.~\cite{troyanovskii1981electron} Edel'man and Faley further explored this system using cyclotron resonance methods, confirming that the effective electron mass is close to the free electron mass and that the electron mobility is significantly impacted by surface defects. They reported a mobility of approximately $8\times 10^4~ T^{-1}$~cm$^2$/Vs in the temperature range between 5~K and 12~K.~\cite{edel1983investigation}
Adams and Paalanen investigated the effects of disorders on the transport properties of a Boltzmann distribution of electrons on solid hydrogen with electron mobility of $0.2-6 \times 10^4$~cm$^2$/Vs.~\cite{adams1987localization} They observed Drude behavior on clean crystals and both weak and strong localization on disordered surfaces.

Kajita and colleagues conducted systematic studies on the electron mobility on solid neon surfaces. They demonstrated that electrons exhibit high mobility on thin helium films adsorbed on the neon surface, governed by scattering mechanisms such as gas-atom scattering and surface-roughness scattering.~\cite{kajita1982two} As the helium film thickness increases, electron mobility decreases, which can be interpreted as the formation of polaron-like states where the electron induces a localized surface deformation.~\cite{kajita1982two,kajita1984new} Kajita {\etal} further examined the stability of electrons on thin helium films adsorbed on solid neon, noting that the strong image force from the substrate leads to deeper bound states compared to bulk liquid helium, which facilitates higher electron densities and stable localization.~\cite{kajita1983stability,kajita1984new} At sufficiently high electron density, they reported the observation of Wigner crystallization of two-dimensional electrons, highlighting the significant role of electron correlation in the transport phenomena at these densities.~\cite{kajita1985wigner} Later, Kono {\etal} studied how adsorbed helium films influence the 2D electron mobility on solid hydrogen.~\cite{kono1991surface1,kono1991surface2} A review was provided by Leiderer, which summarizes experimental and theoretical advancements in understanding surface electron dynamics.~\cite{leiderer1992electrons} Together, these studies underscore the critical influence of surface conditions and electron interactions on mobility, providing a comprehensive understanding of electron dynamics in 2D electronic systems on nonpolar solid substrates.

When many electrons are confined on the surface of a QLS, they can form 2D electron gas, liquid, or solid, depending on the electron number density $n_e$ and temperature $T$.~\cite{platzman1974phase} At a given temperature, each phase can be further divided into the low-density classical-nondegenerate regime and the high-density quantum-degenerate regime. The classical electron gas and liquid are, respectively, called the Coulomb gas and liquid. The quantum electron gas and liquid are, respectively, called the Fermi gas and liquid. Similarly, the classical and quantum solids are known as the classical and quantum Wigner solids, respectively.~\cite{wigner1934interaction} 

All these phases, in both classical and quantum regimes, can be categorized by the so-called plasma parameter $\mathit{\Gamma}(n_e,T)$, which quantifies the competition between the electron-electron interaction and the free-electron kinetics. It reads
\begin{equation}
\mathit{\Gamma}(n_e,T) = \frac{U_e(n_e)}{K_e(n_e,T)},
\end{equation}
where $U_e(n_e)$ and $K_e(n_e,T)$ are, respectively, the mean Coulomb energy per electron and mean kinetic energy per electron.~\cite{platzman1974phase} The phase separation conditions are listed in Table~III, where $\KB$ is the Boltzmann constant, 
\begin{equation}
	\EF(n_e)= \frac{\pi\hbar^2n_e}{m_e}
\end{equation}
is the Fermi energy, proportional to $n_e$ at $T=0$, of a 2D noninteracting Fermi gas, and $\mathit{\Gamma}_0$ is the solid-to-liquid melting-transition parameter, which is on the order of 100. Classical Monte Carlo (CMC) simulation shows $\mathit{\Gamma}_0 \approx127$ for classical melting of a classical Wigner solid,~\cite{gann1979monte,fang1997internal} whereas quantum Monte Carlo (QMC) simulation shows $\mathit{\Gamma}_0\approx 72$ for quantum melting of a quantum Wigner solid as $T\rightarrow 0$.~\cite{drummond2009phase}

\begin{table}[htb]
	\caption{Conditions of gas, liquid, and solid phases of 2D electrons in the classical and quantum regimes.~\cite{platzman1974phase}}
	\centering
	\begin{tabular}{|c|c|c|}
		\hline
		\makecell{} & \makecell{$\EF(n_e) \lesssim \KB T$} & \makecell{$\EF(n_e) \gtrsim \KB T$} \\
		\hline
		\makecell{$\mathit{\Gamma}(n_e,T) \lesssim 1$} & \makecell{Classical\\ Coulomb Gas} & \makecell{Quantum\\ Fermi Gas} \\
		\hline
		\makecell{$1 \lesssim \mathit{\Gamma}(n_e,T) \lesssim \mathit{\Gamma}_0$} & \makecell{Classical\\ Coulomb Liquid} & \makecell{Quantum\\ Fermi Liquid} \\
		\hline
		\makecell{$\mathit{\Gamma}(n_e,T) \gtrsim \mathit{\Gamma}_0$} & \makecell{Classical\\ Wigner Solid} & \makecell{Quantum\\ Wigner Solid} \\
		\hline
	\end{tabular}
	\label{Tab:ElectronicPhases}
\end{table}

To produce a quantitative phase diagram over $(n_e,T)$, the analytical forms of $U_e$ of a classical Coulomb gas and $K_e$ of a quantum Fermi gas are customarily used,~\cite{platzman1974phase} 
\begin{align}
	U_e &= \frac{e^2}{r_e} = e^2\sqrt{\pi n_e}, \\
	K_e &= \int_0^\infty \Dd k~ \frac{\DS\frac{\hbar^2k^3}{2m_e\pi n_e}}{\exp\left[\DS\left(\frac{\hbar^2k^2}{2m_e}-\mu\right)/\KB T\right]+1},
\end{align}
where $r_e$ is the nearest inter-electron distance, which is related to the electron density by $n_e=1/\pi r_e^2$, and
\begin{equation}
	\mu= \KB T\ln[\exp(\EF(n_e)/\KB T)-1]
\end{equation}
is the chemical potential. At high $T$ and low $n_e$, $\EF(n_e)\lesssim\KB T$, $K_e \approx \KB T$, the system behaves classically. The classical melting condition is thus
\begin{equation}
	\mathit{\Gamma}^{\text{classical}} = \frac{e^2}{r_e\KB T} = \frac{e^2\sqrt{\pi n_e}}{\KB T} = \mathit{\Gamma}_0.
\end{equation} 
In contrast, at low $T$ and high $n_e$, $\EF(n_e)\gtrsim\KB T$, $K_e\approx \frac{1}{2}\EF =\pi\hbar^2n_e/2m_e$, the system behaves quantum mechanically. The quantum melting condition, at $T=0$ in particular, can be found as
\begin{equation}
	\label{eq:Gamma}
	\mathit{\Gamma}^{\text{quantum}}(T=0) = \frac{2e^2m_e}{\hbar^2\sqrt{\pi n_e}}=\frac{2r_e}{\aB}  \equiv 2r_s = \mathit{\Gamma}_0,
\end{equation} 
where $\aB=\hbar^2/e^2m_e$ is the standard Bohr radius and $r_s=r_e/\aB$ is the dimensionless inter-particle distance measured in $\aB$. The melting condition is called the Lindemann criterion.~\cite{platzman1974phase}

\begin{figure}[htb]
	\includegraphics[scale=0.9]{\imgpath/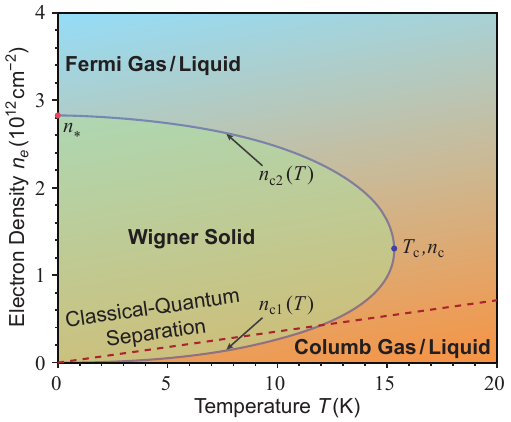}
	\caption{Phase diagram of a 2D electronic system. The orange and blue colored areas represent the classical and quantum gas and liquid phases, respectively. The dome area bounded by the purple curve is the Wigner solid phase. The dashed red line roughly separates the classical and quantum regimes.}
	\label{Fig:WignerPhase}
\end{figure}

Figure~\ref{Fig:WignerPhase} shows the calculated liquid-solid phase boundary by using $\mathit{\Gamma}_0 = 127$. The dashed straight line represents the classical-quantum separation $\EF(n_e)=\KB T$. At a typical experimental temperature of $\leq1$~K, the transition to Wigner crystal occurs at $n_e$ below about $10^9$~cm$^{-2}$ with $\EF (n_e)\ll \KB T$. Therefore, the electrons in this regime obey the classical Boltzmann statistics and the transition resembles a classical melting. In the quantum regime far above the dashed line where $K_e\approx \pi\hbar^2n_e/2m_e$, $K_e$ can outpace $U_e$ as $n_e$ increases. At a critical density at $T=0$,
\begin{equation}
n_{*}\simeq\frac{4e^4m^2_e}{\pi\hbar^4 \mathit{\Gamma}^2_0} = 2.8\times10^{12}~\text{cm}^{-2},
\end{equation}
quantum melting of the Wigner crystal can occur even at zero temperature. At any finite temperature, there are two critical densities $n_{\text{c}1}(T)$ and $n_{\text{c}2}(T)$, roughly corresponding to the classical and quantum melting, respectively. The two melting curves merge at a temperature $T_{\text{c}}\simeq15.3$~K and density $n_{\text{c}}\simeq1.3\times10^{12}~\text{cm}^{-2}$. 

The first experimental proof of Wigner crystallization in the classical regime, $\EF(n_e)\ll \KB T$, is in the electron-on-liquid-helium (eHe) system,~\cite{grimes1979evidence} with the measured $\mathit{\Gamma}_0 \simeq 137$, very close to the theoretical prediction. Extending to higher temperature following the classical melting curve up to $T_\text{c}$ is not possible for eHe, because helium transitions into a gas phase at a much lower temperature. For quantum melting at $T\approx 0$, the criterion $2r_s=\mathit{\Gamma}_0$ gives a quantum phase transition critical density $n_*$. However, the surface instability of liquid He limits the electron density to below $2.4\times10^9$~cm$^{-2}\ll n_*$,~\cite{leiderer1992electrons} preventing observation of quantum melting in the eHe system. Note that the evaluation of the critical density can be affected by the thickness of QLS films. For sufficiently thin films, the screening effect from the (metal or dielectric) substrates should be considered. If the electron-substrate distance is smaller than the inter-electron distance, the screened Coulomb interaction can be described by the Rytova-Keldysh potential.~\cite{rytova2018screened, keldysh2024coulomb}

In order to achieve a high $n_e$ to explore the quantum regime of 2D electrons, efforts have been made by trapping the electrons on a thin helium film on a dielectric substrate~\cite{leiderer1992electrons,gunzler1996evidence} or on liquid He in narrow channels~\cite{bradbury2011efficient,papageorgiou2005counting,kawakami2023blueprint} so the surface instability can be mitigated. There were also attempts to put a thin superfluid He film on top of another cryogenic substrate (such as solid hydrogen) to increase the electron density while keeping a high mobility on the order of $10^5$~cm$^2$/Vs. However, it was found that while the electron density does increase, the mobility decreases more significantly due to the introduced rougher solid surface and the added scatterings with liquid ripplons.

It has also been shown experimentally that solid neon can host a much higher surface electron density above $3\times10^{11}$~cm$^{-2}$. However, the measured mobility was only on the order of $10^3$~cm$^2$V$^{-1}$s$^{-1}$ due to rough surfaces.~\cite{kajita1984new} Besides, atomic disorders on a rough surface tend to localize electrons more strongly than the Coulomb interaction through the mechanism of Anderson localization.~\cite{ahn2023density} Therefore, realization of a genuine Wigner crystal on a quantum solid is still challenging.

\section{Quantum Electronics on Liquid Helium}\label{Sec-eHe}

\subsection{Transport collective electrons on helium}

The mobility of electrons hovering over a liquid He surface was first measured by Sommer and Tanner using an ingeniously simple setup.~\cite{sommerPRL1971tanner} Because it is impossible to achieve a direct electrical contact between such a system and dc leads, the authors used a set of electrodes submerged below the He surface and coupled capacitively to the electrons. By driving one of the electrodes with an ac voltage and detecting a signal coupled by the surface charge to another electrode, the electron mobility could be obtained from the change of phase signal. The Sommer-Tanner (ST) method became a major experimental technique to investigate the electronic properties of quantum liquid systems. As an extremely clean system free of defects and static disorder, the quantum-confined electrons on the surface of liquid He present an ideal playground for the experimental study of the low-dimensional classical many-particle systems. Comparing with other 2D electroic systems realized, for example, in the inversion layer of a semiconductor-insulator interface or in the semiconductor heterojunction, electrons on helium showed the record-high mobility exceeding $10^8$~cm$^2$/Vs.~\cite{shirahamaJLTP1996mobility} A pristine ``soft" liquid substrate, combined with an unscreened Coulomb interaction between electrons, facilitated discovery of many fascinating phenomena such as the Wigner crystallization,~\cite{grimes1979evidence,fisher1979phonon} the Bragg-Cherenkov scattering of an electron solid,~\cite{kristensen1996hall,dykman1997bragg} chiral edge-magnetoplasmons,~\cite{mast1985observation,glattli1985dynamical,peters1991observation,kirichek1995observation} quantum magnetotransport of an electron fluid,~\cite{dykman1993magnetoresistance,lea1994many,monarkha2002quantum} and photo-induced zero-resistance and incompressible states,~\cite{konstantinov2010photon,chepelianskii2015incompressible} to mention a few. However, it has been proved to be very difficult to reach the quantum degeneracy regime in this system,~\cite{etz1984stability,gunzler1996evidence,shikin2001dip} thus limiting the possibility to study the quantum Hall effect and related phenomena.

The development of microscopic electronic devices, such as the metal-oxide-semiconductor field-effect transistor (MOSFET), which marked the second half of the last century, induced an enormous impact on both fundamental sciences and industrial applications. Most of the early research on eHe focused on a macroscopic pool of electrons covering a large area of a bulk liquid. The first microscopic structure with eHe was attempted by Marty who prepared an electron system on a fractionated helium surface.~\cite{marty1986stability} The grooves between the 35~\textmu m wide and 5~\textmu m thick stripes of a copper meandering line was filled with superfluid He by the capillary forces and charged with electrons produced by a glow discharge above the device. The main motivation of this work was to suppress the hydrodynamic instability of the liquid surface. An electron density of $\SI{4.1e9}{cm^{-2}}$ was reached in such a device, although no measurements of the electron mobility was reported. The first functioning Helium-FET was constructed by Klier {\etal} using eHe on a structured metal substrate.~\cite{klier2000first} Rather unusual for eHe, the device was operated in a dc current mode by continuously charging the liquid surface with electrons emitted from a hot tungsten filament placed above the device. Driven by a dc potential difference between the source and drain electrodes, the surface electrons pass through a narrow channel formed by the electrostatic potential from a voltage-biased split-gate electrode, thus realizing 2D or quasi-1D transport of charges in a fashion similar to MOSFET.

\begin{figure}[t]
	\includegraphics[width=\columnwidth,keepaspectratio]{\imgpath/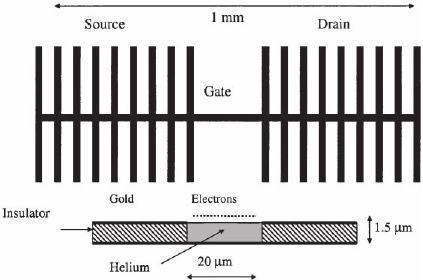}
	\caption{\label{fig:1} Outline of the electron-on-helium FET showing the microchannel array geometry (top-view) and cross-section of a conducting channel. Adapted with permission from Ref.~\cite{glasson2000microelectronics}. Copyright 2000, Elsevier.}
\end{figure}

A disadvantage of the first Helium-FET is the low values of electron mobility 1~cm$^2$/Vs caused by pinning effects due to metal substrate roughness.~\cite{klier2000first} Later devices showed significantly improved electron mobility.~\cite{shaban2016helium} In order to maintain the high values of mobility comparable to that on the bulk helium, a new type of Helium-FET was developed by Glasson {\etal}~\cite{glasson2000microelectronics,glasson2001observation} Similar to the experiment by Marty, the conductance of the device was through the eHe filling $\sim$16--30~\textmu m wide and $\sim$1--2~\textmu m deep channels prepared using photolithography on silicon. The whole device consisted of two arrays of such channels cross-connected by a 1~mm-long single channel. See Fig.~\ref{fig:1}. The FET operation was defined by the gold electrodes at the bottom of the channels, with two channel-arrays and the central connecting channel, which act as the source, drain and gate, respectively. The source and drain current could be measured by the standard ST method and controlled by a dc bias voltage applied to the gate electrode by varying the number of electrons in the conducting channel. Along with high electron mobility $\lesssim\SI{1e5}{cm^2/Vs}$ and electron density $\SI{3.1e9}{cm^{-2}}$, above the hydrodynamic stability limit on bulk helium, this novel device revealed an unusual nonlinear transport of electrons through the gate channel.~\cite{glasson2001observation} The conductance of the device showed an oscillating behavior, which was interpreted as a novel phase of spatially ordered current filaments of electrons aligned along the edge of the gate channel. Later, it was shown that such behavior appears due to the dynamical recoupling between the electron solid in the channel and surface deformation of the liquid substrate.~\cite{zou2021dynamical}

\begin{figure}[b]
	\includegraphics[width=\columnwidth,keepaspectratio]{\imgpath/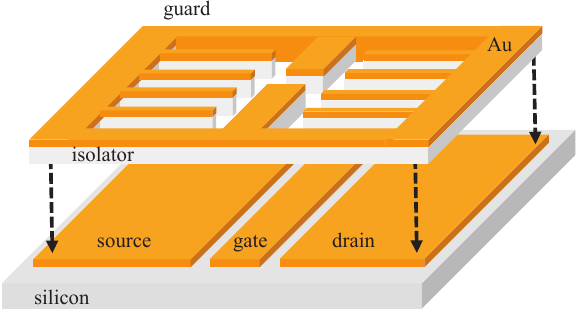}
	\caption{\label{fig:2} Schematic drawing of a fabricated microchannel device for electron transport measurements.}
\end{figure}

Microchannel devices similar to the eHe FET developed by Glasson {\etal} proved to be an extremely valuable tool to investigate the transport properties of surface electrons on superfluid helium. Employment of such devices allowed to observe and study new phenomena in 2D and quasi-1D electronic systems, such as nonlinear transport of Wigner solid,~\cite{ikegami2009nonlinear,ikegami2010melting,badrutdinov2016nonlinear,lin2018sliding} reentrant melting of a quasi-1D Wigner crystal,~\cite{ikegami2012evidence,g2013reentrant} dynamical recoupling (stick-slip) between Wigner solid and liquid helium substrate,~\cite{rees2016stick,zou2021dynamical}, and ripplon-polaron charge transport through a T-junction.~\cite{badrutdinov2020unidirectional} The schematic drawing of a typical microchannel device fabricated on a piece of silicon wafer by photolithography is shown in Fig.~\ref{fig:2}. Two gold layers, separated by an insulating layer of hard-baked photoresist, are patterned by lift-off, forming a set of electrodes that are used to confine and control the electrons. The sub-micron gaps between the source, drain and gate electrodes of the bottom layer can be made by e-beam lithography. The top layer serves as a negatively-biased guard electrode to improve confinement of electrons in the channels and to avoid charging the top of the channels covered by a thin superfluid helium film. A separate pair of electrodes forming a split-gate is sometimes introduced at the top of the gate channel.~\cite{rees2016stick,rees2017bistable} Other materials, such as SiO$_2$, can be used as the insulating layer between bottom and top metal electrodes.~\cite{ikegami2015melting} The typical thickness of the insulating layer, which defines the depth of the channels, varies from half to a few microns, although a device with channel depth as small as 200~nm has been reported.~\cite{asfaw2019transport}

\begin{figure}[t]
	\includegraphics[width=\columnwidth,keepaspectratio]{\imgpath/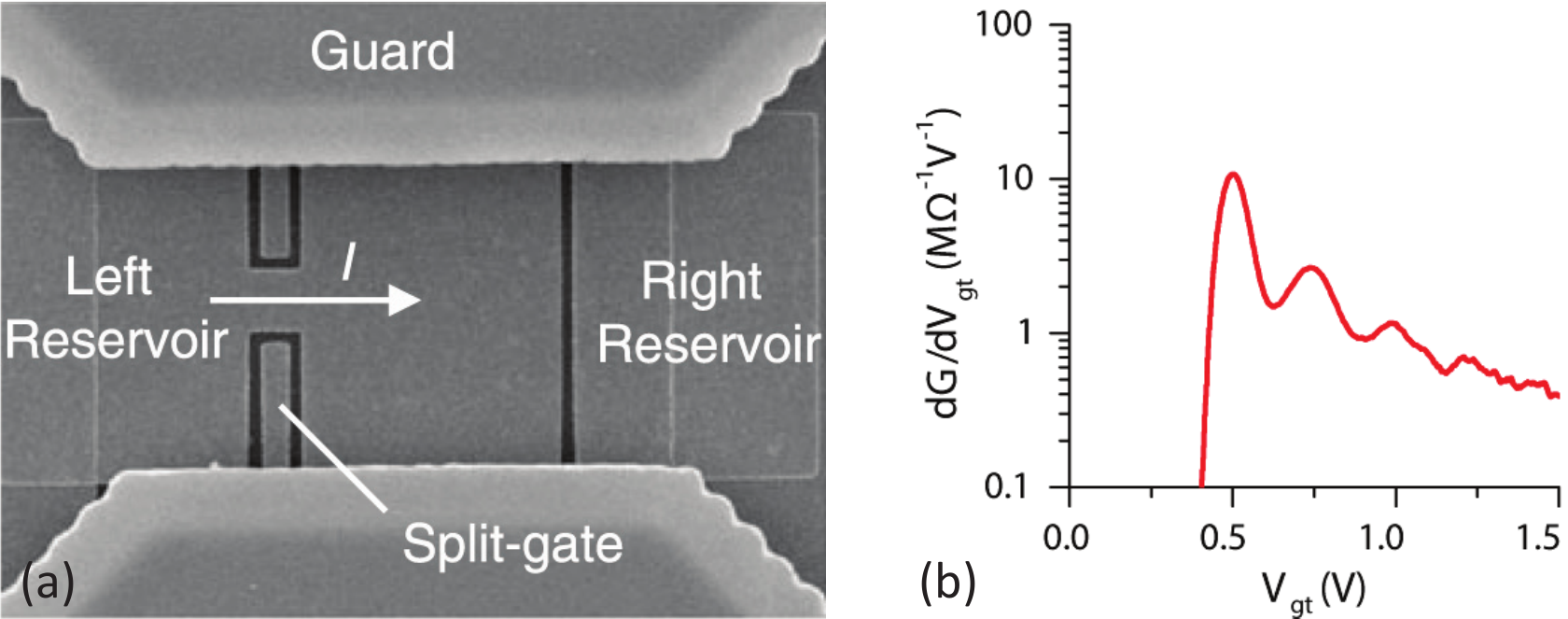}
	\caption{\label{fig:3} Electron-on-helium PC device: (a) SEM image of the device, and (b) differential conductance of the device. Adapted with permission from Ref.~\cite{rees2011point}. Copyright 2011, American Physical Society.}
\end{figure}

The electron density in the conducting channel above the gate electrode can be varied in a wide range from zero to above the hydrodynamic limit on bulk helium, realizing different phases of electronic systems from a dilute gas to a solid. Electron densities approaching $\SI{1e10}{cm^{-2}}$ have been reported.~\cite{leiderer2016stability} However, such a density is still orders of magnitude smaller than that required to reach quantum melting. Therefore, many devices based on the quantum-degenerate 2D electron gas (2DEG) in semiconductors are not possible with eHe. Nevertheless, some advanced devices with electrons confined in microchannels were realized and studied. Rees {\etal} reported a classical analog of a quantum-point-contact (QPC) device.~\cite{rees2011point,rees2012transport,rees2012commensurability} The flow of surface electrons through a 10~\textmu m wide and 20~\textmu m long channel was subject to a constriction formed by a split-gate beneath the helium surface. See Fig.~\ref{fig:3}(a). The current $I$ and conductance $G$ of electrons could be measured by the standard ST method. By varying the bias potential $V_{\rm gt}$ applied to the split-gate, a periodic structure in the measured $I$, $G$ and the differential conductance $dG/dV_{\rm gt}$ has been observed. See Fig.~\ref{fig:3}(b). This behavior was attributed to the effect of the Coulomb repulsion between electrons moving through a constriction, where each peak in $dG/dV_{\rm gt}$ corresponds to an increased number of electrons simultaneously passing through the constriction. Thus, close to the conductance threshold, transport of one electron at a time was realized.

In addition to the microfabricated channel devices, other setups have been employed to achieve a quasi-1D electronic system on helium and study its transport. Kovdrya and Nikolaenko used an optical diffraction grating as a dielectric substrate.~\cite{kovdrya1992quasi} Electrons were confined inside the grating grooves covered by the superfluid He film. Similar method was used by Yayama and Tomokiyo.~\cite{yayama1996anisotropy} In general, the mobility of electrons in such devices was found to be lower than expected, likely due to the effect of a random potential seen by mobile electrons from pinned electrons by the dielectric substrate covered by the thin part of superfluid helium film.~\cite{kovdrya1998mobility}

\subsection{Sense individual electrons on helium}

Significant efforts towards development of electronic devices on an QLS surface were motivated by proposals to employ them as qubits for quantum computing,~\cite{platzman1999quantum,smolyaninov2001electrons} as discussed in details below. An essential requirement for the realization of qubits is to trap, control, and detect individual electrons.
Unlike the trapped charged particles in a vacuum using Penning or Paul traps, trapping surface electrons on a QLS surface only requires in-plane electrostatic fields, because electrons naturally form bound states in the out-of-plane direction. The in-plane trapping potential can be easily realized by patterned electrodes close to the QLS surface. 

\begin{figure}[t]
	\includegraphics[width=\columnwidth,keepaspectratio]{\imgpath/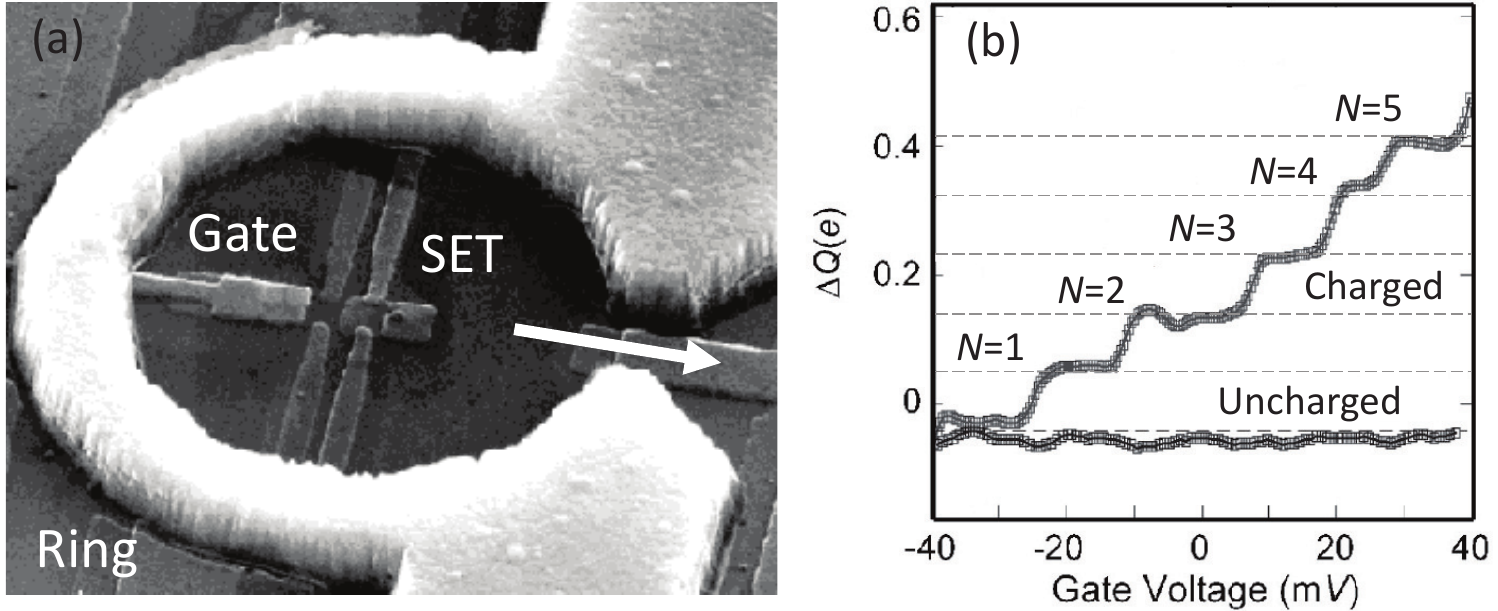}
	\caption{\label{fig:4} Single-electron counting on liquid helium surface: (a) Micrograph of the electron trap and SET device. (b) The reduced image charge at SET (in units of the elementary charge $e$) induced by individual surface electrons in the trap. Adapted with permission from Ref.~\cite{papageorgiou2005counting}. Copyright 2005, AIP Publishing LLC.}
\end{figure}

Detection of individually trapped electrons presents a big challenge. Papageorgiou {\etal} built a setup to manipulate and detect individual electrons by using an aluminum-based superconducting single-electron transistor (SET).~\cite{papageorgiou2003detecting,papageorgiou2005counting} In their experiment, an aluminum ring of an inner diameter 5~\textmu m defines an electron trap filled by 0.8~\textmu m-deep helium. See Fig.~\ref{fig:4}(a). A SET was positioned near the center of the trap beneath the surface, which acts as a sensitive electrometer that detects the image charge induced by the trapped electrons. By varying the dc bias potentials applied to the SET and surrounding electrodes, a variable number $N$ of surface electrons could be trapped above SET and induce image charges in the SET island.  In the experiment, the electron reservoir is a long 10~\textmu m wide channel (not shown) and electrons could be transferred between the reservoir and trap (white arrow) by adjusting the trapping potential with a dc voltage applied to the gate electrode of SET. Fig.~\ref{fig:4}(b) shows the steps in the image charge detected by SET as five electrons leave the trap one by one by decreasing the gate voltage, until the trap is empty. For comparison, the background detection by SET for an uncharged trap is also shown, demonstrating the long-term charge stability of the SET of about 0.01$e$.

Besides direct application of the SET device for qubit manipulation, it can be used to study strongly-correlated few-body systems. Glasson {\etal} pointed out that by analyzing the charging spectra similar to that shown in Fig.~\ref{fig:4}(b), one can obtain information about different structural arrangements of electrons in the trap governed by the competition between the Coulomb repulsion between electrons and their confinement by the trapping potential.~\cite{glasson2005trapping} Rousseau {\etal} used a device similar to the one described above to obtain the addition spectra of $N\leq 20$ electrons confined in a trap.~\cite{rousseau2007trapping,rousseau2009addition} The energies to extract a single electron from an $N$-particle system was obtained from the charging spectra and compared with the results of Monte Carlo simulations. The comparison revealed a variety of ordered ground states of a few-electron system  called Wigner islands, whose structures are different from the triangular lattices of bulk Wigner crystals.

\subsection{Transfer individual electrons on helium}

\begin{figure}[htb]
	\includegraphics[width=\columnwidth,keepaspectratio]{\imgpath/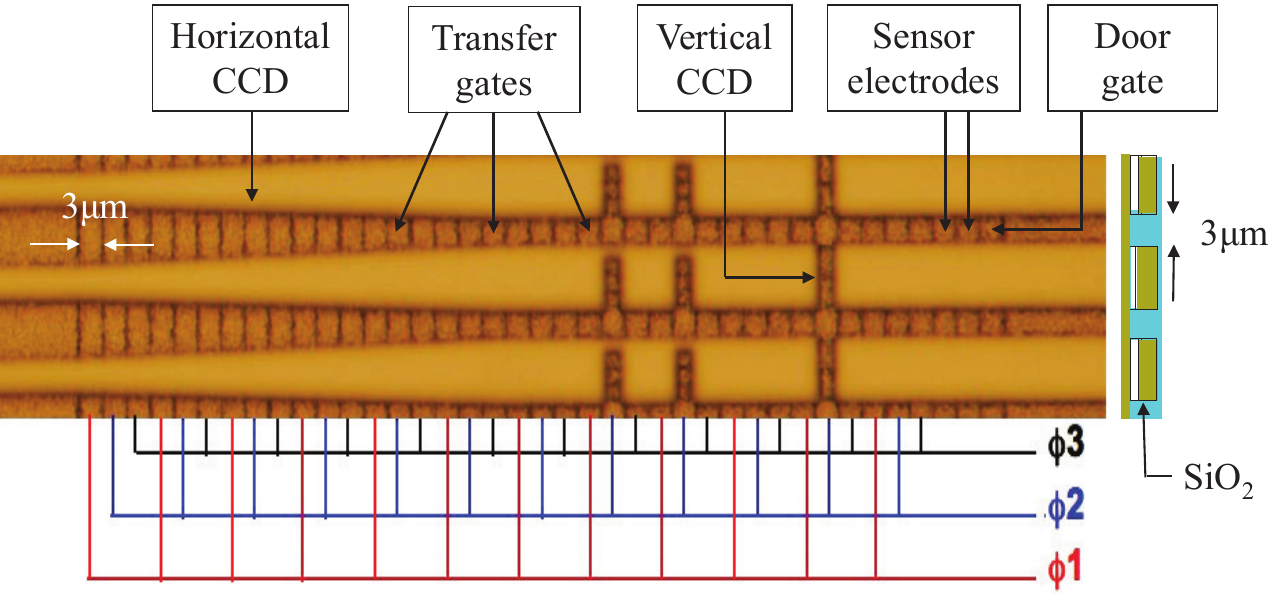}
	\caption{\label{fig:5} Micrograph of a 3-phase multichannel CCD for efficient clocked electron transport on superfluid helium. Three of 120 parallel channels are shown, with a schematic cross section to the right of the image. Adapted with permission from Ref.~\cite{bradbury2011efficient}. Copyright 2011, American Physical Society.}
\end{figure}

A scheme to manipulate a large number of qubits is necessary toward a fault-tolerant quantum computer. The extremely high mobility of electrons on liquid helium brings an advantage in building a scalable quantum computing architecture. Lyon envisioned a quantum computing scheme that incorporates electron spins with a charge-coupled device (CCD), which is well known in semiconductor physics.~\cite{lyon2006spin} In such an arrangement, the mobile electrons can be rapidly moved between different areas of the device while preserving their spin coherence. This potentially allows for a massively parallel quantum gate operation in a large-scale quantum computer.~\cite{kielpinski2002architecture} To demonstrate efficient clocked transfer of electrons on liquid helium, a multichannel helium CCD was developed.~\cite{bradbury2011efficient,sabouret2008signal} In this device, electrons are transferred along gate-defined paths by applying a standard clock voltage sequence to the gate electrodes. The top layer of the device fabricated by CMOS technology was comprised of 120 parallel channels filled with 2~\textmu m deep superfluid helium. See Fig.~\ref{fig:5}. Perpendicular gate electrodes running under all 120 channels had a 3~\textmu m period (including a 0.5~\textmu m gap) and were arranged as a 3-phase horizontal CCD, with three sets of adjacent gates making up a pixel for electron transfer simultaneously along all 120 channels. A packet of electrons could be controllably loaded into the device from an electron storage (on the right side in Fig.~\ref{fig:5}) by lowering the potential barrier from a voltage-biased door gate and detected by means of two sensor electrodes using the standard ST method. Then, this packet of electrons was loaded into the rightmost pixel of the transfer gates and clocked along the channels at the rate of 240~kHz by a programmed 3-phase clock voltage sequence on the gates. Electrons could be moved back to the sensor electrodes for charge detection after any number of clocked cycles, thus providing information about the transfer efficiency of the device. It was found that no detectable loss of charges occurs during the transfer across $10^9$ pixels in total (moving electrons 9~km), regardless the size of the electron packet. Such an unprecedented efficiency of the helium CCD by far surpass any conventional semiconductor devices, owing to the high mobility of electrons on helium and strong fringing fields across the microchannel CCD gates.

In addition to the multichannel horizontal CCD, the device featured a single perpendicular channel with underlying gates forming a vertical CCD (see Fig.~\ref{fig:5}). This CCD allowed to transfer electrons between different channels of the horizontal CCD. Bradbury {\etal} used a clocking sequence where electrons could be shifted by one pixel left or right along the channels, with vertical interchannel transfer of electrons in between the horizontal shifts, to emulate 2D transport of electrons in the device.~\cite{bradbury2011efficient} This transport demonstrated the same high efficiency as the transport in the horizontal CCD, thus showing that complex and parallel operations on many qubits can be realized for the purpose of large-scale quantum computing.

In the aforementioned device, the number of electrons in the packet varied from a few electrons per channel to less than one electron on average (meaning some channels were empty). However, for robust scalable operations it is desirable to eliminate this transfer uncertainty. For this purpose, Takita and Lyon introduced an electron turnstile for each channel of CCD that allowed to deplete packets of electrons in each channel in a controllable way.~\cite{takita2014isolating} It consisted of a narrowed 0.8~\textmu m portion of channel with five gate electrodes creating a controllable asymmetric double-potential well at the helium surface. A packet of electrons brought to the turnstile region was sequentially split between two potential wells until all remaining electrons in the packet resided in one of the wells. After the depletion sequence, the electrons from all the 78 parallel channels of the device were detected by the procedure described earlier. It is expected that the minimum number of electrons in the depleted packet must be equal to one. The experimental results indeed showed saturation of the electron signal with depletion, with a fixed average number of electrons per channel. However, signal calibration indicated that the signal approximately corresponds to two electrons per channel, rather than one electron as expected. It was concluded that more accurate measurements are required. Nonetheless, the device demonstrated a reliable way to produce and transport quantized charge on a superfluid helium surface.

The CCD scheme above showcases the benefits of using mesoscopic devices developed for semiconductors to manipulate electrons floating on liquid helium. Other type of devices and methods from other areas can be potentially useful for such purpose as well. Recently, Byeon {\etal} achieved the coupling between floating electrons and piezoelectric surface acoustic waves (SAW) and demonstrated acoustoelectric transport in such a system.~\cite{byeon2021piezoacoustics} In their device, electrons are held on a 70~nm thin superfluid film covering a highly-polished surface of the lithium niobate piezoelectric substrate. Travelling SAW are excited in the substrate by an interdigitated transducer (IDT), thus producing an evanescent electric field near the surface that couples to the electrons. This produces traveling charge density waves of surface electrons that could be detected by capacitive coupling to an electrode at the bottom of the substrate. This first demonstration of the acoustoelectric transport in electrons on helium provides a novel toolkit for their control. Combined with microchannel and SET devices, it can be potentially used in various applications, such as flying electron qubits.

\subsection{Electron Rydberg-state and spin qubits on helium}

The most appealing application of eHe is perhaps using each electron as a quantum bit for quantum information processing. The necessity to control and read out the quantum states of a single electron presents a new challenge for the field. If this challenge can be overcome, eHe promises a scalable quantum platform with qubits above an ultraclean substrate of a quantum liquid. The first proposal was made by Platzman and Dykman who suggested to use the quantized out-of-plane motion of electrons (two lowest Rydberg states) as the qubit states.~\cite{platzman1999quantum} The advantage of using the Rydberg states is the long-range Coulomb interaction between electrons. Since the mean distance between two electrons depends on their state occupation, the Coulomb repulsion results in a state-dependent interaction energy which can be used to entangle two qubits. For example, for two ground-state electrons localized in the plane at a distance 1~\textmu m apart, their interaction energy changes by about 100~MHz when one of the electrons is excited. This introduces a similar order-of-magnitude shift of the Rydberg transition frequency of a qubit conditioned on the state of the neighbour qubit, which allows to implement a controlled-NOT two-qubit gate. It was also pointed out that, by localizing a single electron in an electrostatic trap that quantizes its lateral motion, the decay of the excited qubit state due to the quasi-elastic one-ripplon scattering can be suppressed, thus promising a long coherence time of such qubits.~\cite{dahm2002using,dykman2003qubits} However, later it was argued that the decay time of such qubits cannot be made longer than approximately 1~\textmu s due to the spontaneous emission of a pair of short-wavelength ripplons.~\cite{monarkha2010decay} This prediction has been confirmed in a recent experiment with a many-electron system on bulk helium.~\cite{kawakami2021relaxation} Such relatively short relaxation of the qubit state imposes a significant constrain on the fidelity of the quantum logic gates, thus making the Rydberg-based qubits be a less attractive candidate for a scalable quantum computer.

Lyon proposed to use the spin of electrons above the surface of liquid $^4$He as qubits.~\cite{lyon2006spin} Residing in vacuum relatively far ($\sim\SI{10}{nm}$) from the surface, such qubits are negligibly affected by the nuclear spin of $^3$He atoms, the only impurity atoms floating on the superfluid helium surface. Comparing with 2DEG in semiconductors, such as silicon and Si/SiGe heterostructures, the intrinsic spin-orbit interaction of surface electrons on helium is orders of magnitude smaller, which implies a spin coherence time exceeding hundreds of seconds. One disadvantage of using the spin of electrons on liquid helium is a very weak magnetic dipole interaction between them, which for two electrons separated by a distance of 1~\textmu m is only of the order 1~Hz. Another disadvantage is the lack of any reliable methods for the spin-state detection in this system. Owing to the small coupling between the magnetic dipole and cavity modes, the spin sensitivity in traditional electron spin resonance (ESR) techniques is significantly constrained.

It was pointed out that using electronic dipole spin resonance (EDSR) could be advantageous. Schuster {\etal} proposed to couple the spin of an electron trapped laterally  on the liquid helium surface to the states of its quantized in-plane motion by introducing a local magnetic field gradient from a current passing through superconducting wire.~\cite{schuster2010proposal} This proposal will be elaborated in a later section. 

\subsection{Hybrid charge-spin qubits on helium}

\begin{figure}[htb]
	\includegraphics[width=\columnwidth,keepaspectratio]{\imgpath/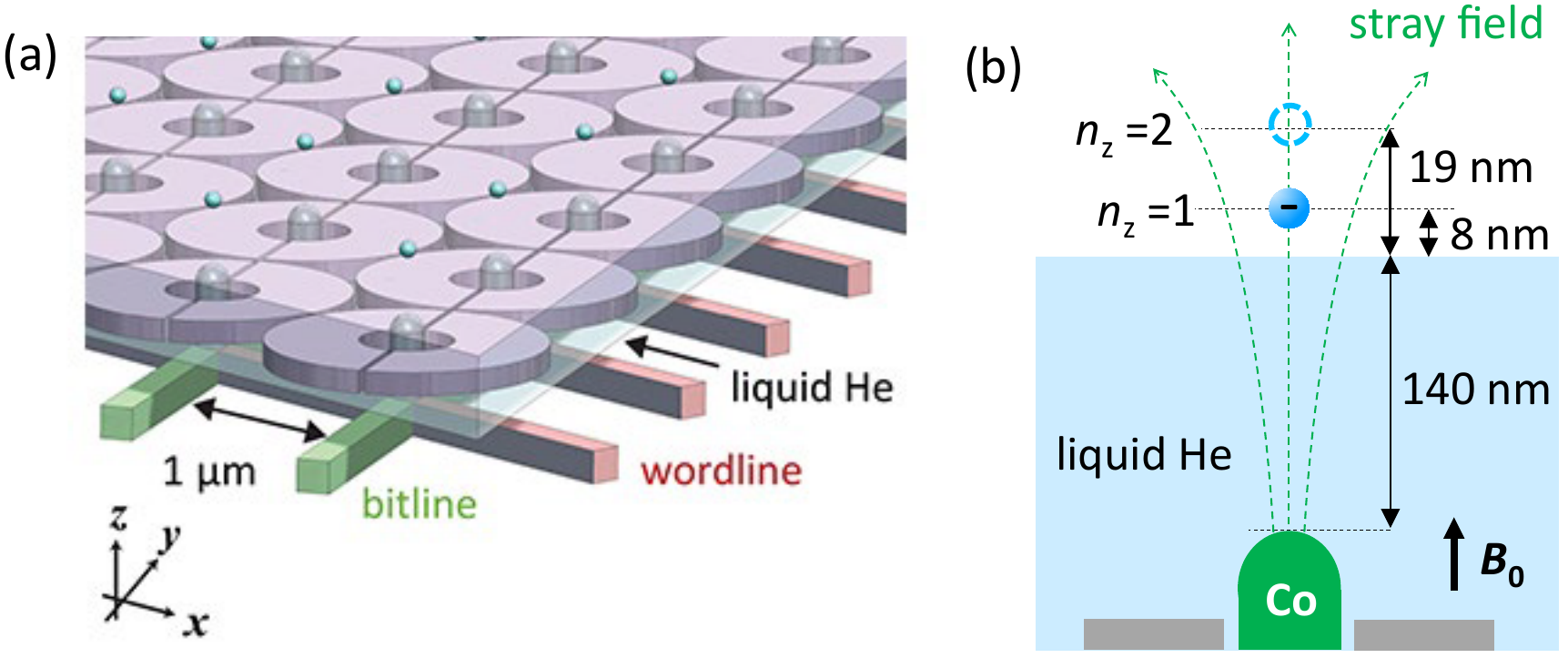}
	\caption{\label{fig:6} A hybrid Rydberg-spin qubit architecture: (a) Electrons (blue circles) are electrostatically trapped over a 2D array of nanofabricated micromagents. (b) Different magnitude of the stray magnetic field on the electrons occupying different Rydberg states with the quantum number $n_z$. Adapted with permission from Refs.~\cite{kawakami2023blueprint,jennings2024quantum}. Copyright 2023 and 2024, American Physical Society.}
\end{figure}

It is very attractive to exploit both the long coherence of spin states of eHe and the large interaction energy of their Rydberg states to create a scalable architecture of high-fidelity quantum gates. Such a hybrid approach was suggested by Kawakami {\etal}~\cite{kawakami2023blueprint} who proposed to couple spins of trapped eHe to their orbital states by a sufficiently strong gradient of the magnetic field in a 2D array of nanofabricated magnetized traps, see Fig.~\ref{fig:6}(a). In such a setup, the Coulomb repulsion between electrons facilitates electrostatic trapping of individual electrons at each node of the array, given that the array geometry is commensurate with the triangular lattice structure of the Wigner solid. Such an architecture allows for parallel addressing of qubits via world lines and bit lines, thus facilitating operations on a very large number of qubits and providing a route towards scalability.~\cite{veldhorst2017silicon,jennings2024quantum} 

The spin state of a trapped electron can be addressed by an electric field from an ac voltage applied to the trapping electrodes in a EDSR manner, thanks to the in-plane gradient of the stray magnetic field from a magnetized cobalt pillar (a micromagnet) at the center of the trap, see Fig.~\ref{fig:6}(b). Physically, the modulation of the electron's in-plane position due to the applied ac field results in an effective ac magnetic field that rotates the spin. The corresponding Rabi frequency was calculated by taking into account the virtual transitions between the lower-energy in-plane orbital states of electron accompanied by the flips of its spin.~\cite{kawakami2023blueprint} The same second-order processes dominate the relaxation of the spin-qubit state, thus imposing a constrain on the fidelity of the single-qubit gate. It was estimated that for an in-plane magnetic field gradient of the order $\SI{0.1}{mT/nm}$ the Rabi frequency of 100~MHz and the spin relaxation time of 50~ms is possible, which potentially results in a very high fidelity of a single-qubit gate exceeding 99.9999\%.~\cite{kawakami2023blueprint}                   
The coupling between two spins of electrons in adjacent traps is possible thanks to the vertical gradient of the stray magnetic field, which couples the spin of each electron to its Rydberg states, and the state-dependent interaction between electrons due to the Coulomb repulsion.~\cite{kawakami2023blueprint} Since the mean distance of an electron from the liquid surface depends on the Rydberg state quantum number, see Fig.~\ref{fig:6}(b), such an electron experiences a different stray magnetic field, therefore different Zeeman splitting of its spin states. This allows to rotate the Rydberg state of each electron spin-selectively using the resonant microwave radiation in a frequency range around 200~GHz (the millimeter-waves). On the other hand, thanks to the Coulomb repulsion between electrons that causes the state-dependent shifts of the Rydberg transition frequencies for each electron, the Rydberg-state rotation of one electron depends not only on its spin state but also on the spin-selectively excited Rydberg state of its neighbour. By applying one $\pi$-pulse and one $2\pi$-pulse of the millimeter-waves to the control qubit and the target qubit, respectively, to rotate their Rydberg states spin-selectively, followed by another $\pi$-pulse to the target qubit to return the system to its initial state, a controlled-phase two-qubit gate can be realized in a manner similar to the Cirac-Zoller gate used for cold trapped ions.~\cite{cirac1995quantum} A disadvantage of such gate is that it suffers from the relatively short relaxation time of the excited Rydberg state ($\sim 1$~\textmu s), which limits the gate fidelity to about 99\%.~\cite{kawakami2023blueprint}

Finally, by virtue of the spin-selective excitation of the Rydberg transition, a quantum-nondemolition (QND) readout of the spin qubit is possible.~\cite{kawakami2023blueprint} In order to separate the Rydberg transition energies for two orientation of spin, the difference of their Zeeman splitting must exceed the Rydberg transition linewidth, which is expected to be in a range 1-10~MHz. For a typical difference of 10~nm between the vertical position of an electron occupying $n_z=1$ and $n_z=2$ states, see Fig.~\ref{fig:6}(b), this requires a vertical gradient of the stray magnetic field $\gtrsim\SI{0.03}{mT/nm}$. Using the millimeter-waves tuned in resonance with the Rydberg transition corresponding to one orientation of qubit's spin, the probability to excite such a transition is high for the chosen orientation of spin, and is negligible for the opposite orientation. Thus, by observing the Rydberg transition of a qubit, its spin state can be detected without affecting it.  

\begin{figure}[htb]
	\includegraphics[scale=0.7]{\imgpath/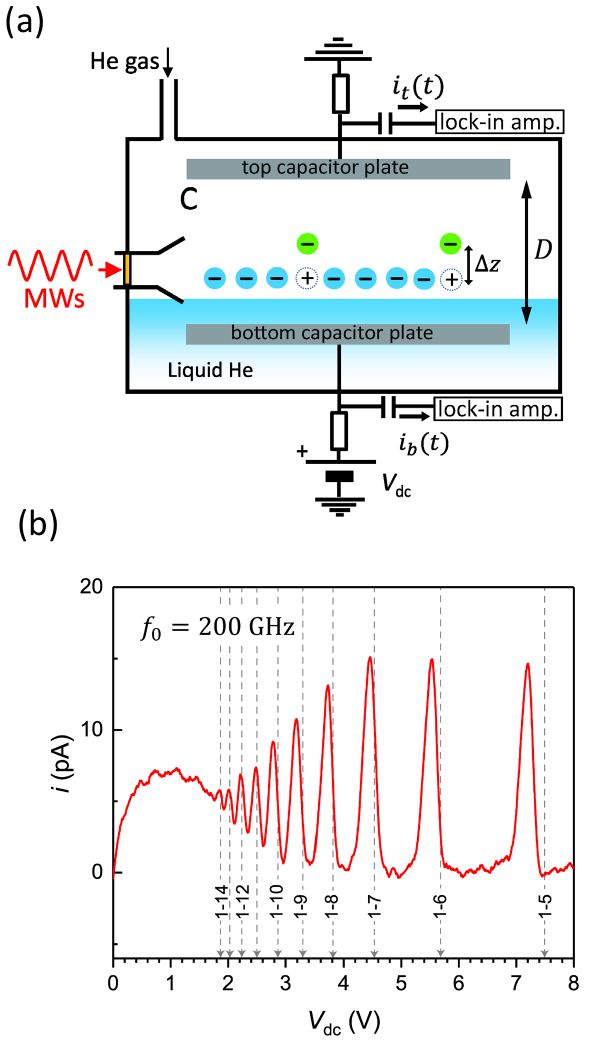}
	\caption{\label{fig:7} Image-charge detection of the Rydberg transition in electrons on helium: (a) the Rydberg transitions of electrons due to incoming microwaves (MWs) induce detectable currents of image charges in the electrodes; (b) the Rydberg spectra of electrons due to their transitions to higher excited states measured by the image current. Adapted with permission from Ref.~\cite{kawakami2019image}. Copyright 2019, American Physical Society.}
\end{figure}     

To detect the Rydberg transition of a single trapped electron, a new method of the image-charge detection was proposed and demonstrated in a many-electron system by Kawakami {\etal}.~\cite{kawakami2019image} The Rydberg transition of an electron causes a change in the image charge induced by the electron in an electrode placed in its proximity. This induces an image current in an electrical circuit connected to the electrode, which can be detected using some sensitive electronics. In the experiment done by Kawakami {\etal}, a large number of electrons on the order of $10^8$ were contained between two parallel plates of a capacitor separated by a distance $D=\SI{2}{mm}$, see Fig.~\ref{fig:7}(a). Electrons were excited by the pulsed-modulated ($\sim\SI{100}{kHz}$) millimeter-wave radiation at the carrier frequency 200~GHz and the demodulated image current was detected by an ordinary lock-in amplifier, thanks to a very large number of electrons that induced an image current on the order 10~pA. The Rydberg transition frequency could be easily tuned in resonance with the carrier frequency of radiation via the Stark shift by adjusting a biasing voltage $V_\textrm{dc}$ at the capacitor's bottom plate. Fig.~\ref{fig:7}(b) shows a typical Rydberg spectra detected by the image current showing a series of the Rydberg transitions of electrons from the ground state to the higher excited states up to the quantum number $n_z=14$. In a following experiment by Kawakami {\etal},~\cite{kawakami2021relaxation} a time-resolved image-current signal due to a pulse-modulated excitation of electrons was detected using a cryogenic two-stage broadband (0.01-100~MHz) amplifier based on a low-noise heterojunction bipolar transistor (HBT).~\cite{elarabi2021cryogenic} This experiment allowed a direct observation of the relaxation of the excited Rydberg states, thus confirming that the relaxation time is limited to about 1~\textmu s by the spontaneous emission of two ripplons. 

\begin{figure}[htb]
	\includegraphics[scale=0.6]{\imgpath/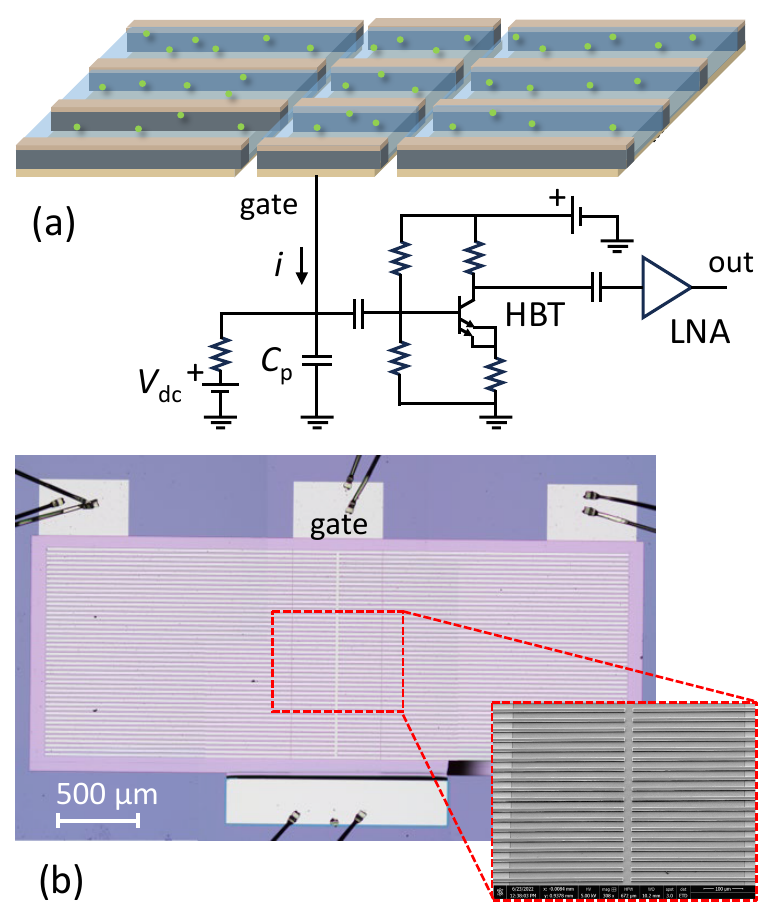}
	\caption{\label{fig:8} Image-charge detection of the Rydberg transition of electrons in microchannels: (a) the image current $i$ induced in the gate electrodes is detected by a two-stage cryogenic amplifier; (b) the microchannel device for electron confinement. Adapted with permission from Ref.~\cite{zou2022image}. Copyright 2022, IOP Publishing.}
\end{figure}

In order to apply the above method for quantum computing with hybrid charge-spin qubits, it has to be scaled down to the detection of the Rydberg transition of a single electron. Zou and Konstantinov pointed out that the image-current signal can be significantly enhanced by bringing electrons much closer to the detection electrodes.~\cite{zou2022image} The image charge difference $\delta q$ induced by the excitation of a single electron in one of two electrodes scales with the distance $D$ between electrodes as $\Delta z/D$, where $\Delta z\approx 10$~nm is the difference between the vertical position of an electron occupying $n_z=1$ and $n_z=2$ states, see Fig.~\ref{fig:7}(a). In the experimental setup employed by Kawakami {\etal},~\cite{kawakami2019image} this image charge difference is on the order $10^{-5}e$. In the experiment by Zou and Kostantinov,~\cite{zou2022image} electrons were confined in an array of 20~\textmu m-wide and 4~\textmu m-deep channels filled with superfluid helium (see Fig.~\ref{fig:8}), similar to the microchannel devices described earlier. Employing such a setup allowed to increase the magnitude of $\delta q$ by two-three orders of magnitude, while reducing the number of electrons to approximately $10^5$. The image current $i$ due to the Rydberg transition of electrons excited by the pulse-modulated (100~kHz) millimeter-wave radiation was detected at the gate electrode at the bottom of the channel array using a cryogenic two-stage amplifier, see Fig.~\ref{fig:8}(a). The current signal was measured as a corresponding voltage drop across a parasitic capacitance $\sim$20~pF of a cryogenic cable connecting the gate to the first-stage HBT preamplifier that served as an impedance-matching network for the 50~$\Omega$ input of a low-noise amplifier (LNA) located at 4~K.~\cite{elarabi2021cryogenic} The observed Rydberg spectra showed a large inhomogenious broadening ($\sim$10~GHz) due to a nonuniform dc electric field experienced by electrons in the microchannels. At the same time, the transition frequency was highly controllable by the dc bias voltages applied to the electrodes of the device. This work demonstrated that the microchannel devices can provide a suitable platform for further work towards quantum-state detection and hybrid charge-spin qubit implementation with eHe.

Two main approaches were suggested to increase sensitivity of the image-charge method towards the level of a single-electron detection.~\cite{zou2022image,kawakami2023blueprint} One approach is to use a high-impedance superconducting resonator as a trans-conductance amplifier to convert a small image current ($\sim$5~fA) induced by the Rydberg transition of a single electron trapped in a microchannel into a voltage signal, with further amplification using a cryogenic low-noise transistor circuit. Such a technique is successfully used to detect the oscillating motion of a single ion in the Penning and Paul traps.~\cite{wineland1975principles,kotler2017hybrid} The superconducting resonator employed in this method is essentially a high-$Q$ parallel $LCR$ circuit that has a large real impedance $R=Q\omega_\textrm{res}L$ at the resonant frequency of the circuit $\omega_\textrm{res}$. Thus, it presents a large load impedance for the image current generated by an oscillating charge, providing that the frequency of charge oscillations coincides with $\omega_\textrm{res}$. For the detection of the Rydberg transition of an electron on helium, modulation of the millimeter-wave excitation must be used to modulate the measured image-current at the frequency $\omega_\textrm{ref}$. To measure such current, a cryogenic resonant amplifier consisting of a helical resonator ($\omega_\textrm{res}=\SI{1.219}{MHz}$ and the loaded quality factor $Q=360$) and a cryogenic high-electron-mobility transistor (HEMT) has been recently developed.~\cite{belianchikov2024cryogenic} With the load impedance of $R=\SI{2.55}{M\Omega}$ and the trans-conductance gain of $\SI{3.2}{nA/V}$, the amplifier demonstrated measured voltage and current noise level of $\SI{0.6}{nV/\sqrt{Hz}}$ and $\SI{1.5}{nA/\sqrt{Hz}}$, respectively, thus making feasible the detection of the Rydberg transition of a single electron with the signal-to-noise ratio SNR=8 and with the measurement bandwidth 1~Hz.~\cite{belianchikov2024cryogenic} 

The second approach towards enhancing sensitivity of the image-charge method is to detect small changes in the resonant properties of a rf (0.1-1~GHz) lumped-element $LC$ circuit coupled to the electrons when they undergo transitions between Rydberg states.~\cite{kawakami2023blueprint,jennings2024quantum} Such a dispersive readout technique has been developed for the detection of quantum transitions in mesoscopic solid-state devices and semiconductor quantum dots.~\cite{vigneau2023probing} Such quantum transitions can cause both resistive and reactive changes in the resonant circuit impedance, which can be detected with a high precision by the rf reflectometry method. In particular, the charge sensitivity as high as $\SI{1.3}{\mu e/\sqrt{Hz}}$ has been recently achieved with this method.~\cite{ahmed2018radio} In case of an electron on helium, it was predicted that by trapping a single electron above an electrode at a distance 140~nm, see Fig.~\ref{fig:6}(b), an image charge difference of $\delta q\sim 0.01e$ is induced in the electrode when electron is excited to the first excited Rydberg state.~\cite{kawakami2023blueprint} With the capacitance sensitivity achieved using the state-of-the-art rf reflectometry, this would allow to detect the Rydberg transition of a single electron with a large measurement bandwidth necessary for a fast qubit-state readout. 

\subsection{Electron charge qubits on helium via circuit QED}

Since 2004, circuit quantum electrodynamics (cQED) based on the interplay of quantized microwave photons on a low-loss superconducting chip and various quantum information systems has gained an increasing popularity.~\cite{wallraff2004strong,blais2004cavity,blais2007quantum} Initially, cQED was utilized mainly to control, readout, and link superconducting Josephson-junction (JJ) qubits for the application of quantum computing.~\cite{blais2007quantum} Later, it was generalized to couple with semiconductor quantum-dot (QD) qubits, molecules, dopants, color centers, rare-earth ions, magnons, phonons, {\etc}, with extended applications into quantum sensing, transduction, and networking.~\cite{burkard2023semiconductor,clerk2020hybrid,blais2021circuit} 

For a typical cQED chip, microwave photons of 2--18~GHz (the S, C, X, and Ku bands) are transmitted through or confined within planar waveguides and resonators. These waveguides and resonators are fabricated on lossless superconducting thin films grown on low-loss dielectric substrates. Common superconducting thin films include aluminum (Al), niobium (Nb), and high-kinetic-inductance (hKI) nitrides, such as titanium nitride (TiN), niobium nitride (NbN), and niobium-titanium nitride (NbTiN). In almost all cases, the substrates are either intrinsic silicon (Si) or sapphire with a loss tangent $\delta<10^{-6}$. Along the waveguides and resonators, microwave photons are concentrated in the gap regions between metal lines and in a close vicinity around the superconducting thin films.

\begin{figure}[htb]
	\centering
	\includegraphics[scale=0.95]{\imgpath/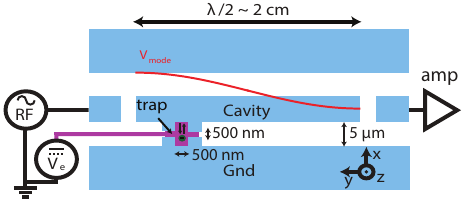}
	\caption{Schematic design of an electron trap on a microwave device. The center stripline and ground planes provide two-dimensional confinement. A dc voltage is provided by a wire insulated from the resonator. Control and readout are performed by microwave input and output. The output signal is amplified by a cryogenic amplifier. Adapted with permission from Ref.~\cite{schuster2010proposal}. Copyright 2010, American Physical Society.}
	\label{Fig:HeCQED}
\end{figure}

The cQED architecture is naturally compatible with on-chip microfluidics, as envisioned by Schuster {\etal} in 2010.~\cite{schuster2010proposal} As shown in Fig.~\ref{Fig:HeCQED} the gap regions between superconducting lines and planes are typically 1--5~{\textmu}m wide and can simultaneously serve as microfluidic channels that host superfluid He with floating electrons on top. If an electron is trapped on the surface of He film with desired thickness, and at the right position in the channel, it can strongly couple with, and be manipulated by, microwave photons in the channel, like a superconducting JJ or semiconductor QD qubit. 

An electron charge qubit in this system can utilize the in-plane motional (charge) states of the electron to encode quantum information. The electric dipole moment of the electron is coupled with the electric field of microwave photons. If the transition frequency of the electron between the ground state and the 1st excited state is at $\sim$6~GHz (in the 4--8~GHz microwave C band with the best cryogenic amplifiers today), then the characteristic size of the electron's wavefunction is on the order of 100~nm (the electron's effective mass in this system is nearly identical to its bare mass). Depending on the type and design of the superconducting resonator, which determines the electric field profile of a single photon in the resonator, the coupling strength $g$ (vacuum Rabi splitting) between the electron and a photon can be estimated to be on the order of 10~MHz.~\cite{yang2016coupling,koolstra2019coupling} A typical photon decay rate $\kappa$ of a resonator is on the order of $0.1$~MHz, dominated by the engineered input-output coupling strength rather than intrinsic loss. Therefore, so long as the qubit linewidth $\gamma$, equivalent to the charge decoherence rate $1/T_2^*$, can be less than $g$, this system can reach the strong coupling regime $g>\kappa,\gamma$ and microwave photons can be used to coherently operate and read out the qubit in the dispersive regime.

\begin{figure}[hbt]
	\centering
	\includegraphics[scale=0.95]{\imgpath/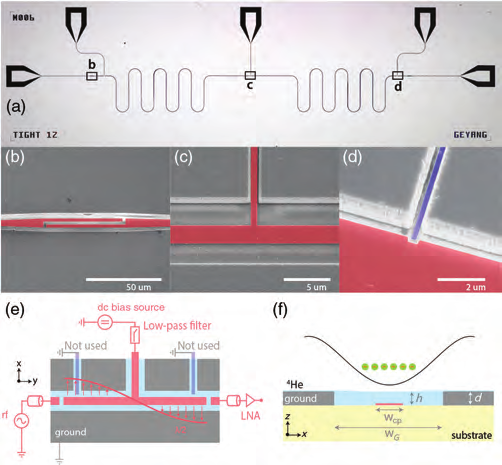}
	\caption{Device design and component configurations. (a-d) Optical
		and scanning-electron-microscopy (SEM) images of a resonator-electron ensemble trap on a superconducting chip. (e) Circuit diagram. (f) Cross-sectional view of the waveguide gap and channel with filled superfluid He and floating electrons. Adapted with permission from Ref.~\cite{yang2016coupling}. Copyright 2016, American Physical Society.}
	\label{Fig:GeDevice}
\end{figure}

\begin{figure}[htb]
	\centering
	\includegraphics[scale=1]{\imgpath/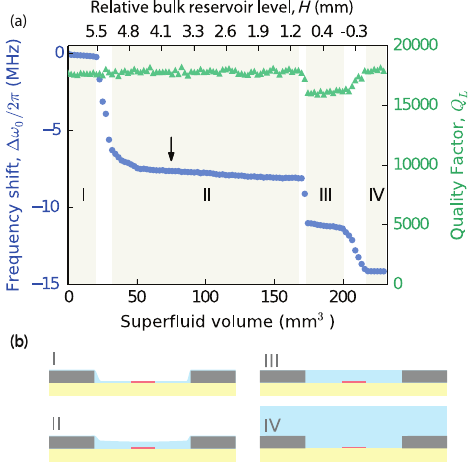}
	\caption{Resonator response to the filling process of superfluid He. (a) Measured
		resonance frequency shift and loaded quality factor change in response to superfluid volume supplied to the cell and relative bulk helium level in the reservoir pit. (b) Different filling state corresponding to the different regimes in (a). Adapted with permission from Ref.~\cite{yang2016coupling}. Copyright 2016, American Physical Society.}
	\label{Fig:GeHeFilling}
\end{figure}

\begin{figure*}[thb]
	\centering
	\includegraphics[scale=1]{\imgpath/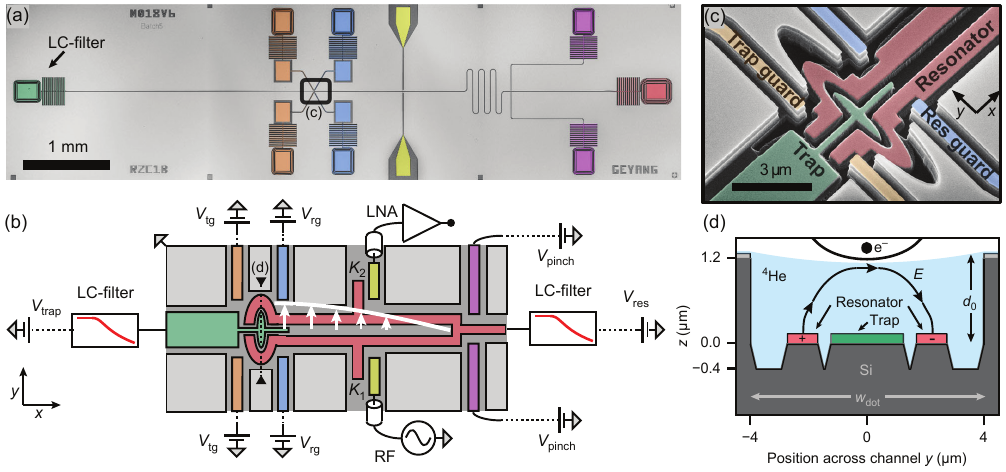}
	\caption{Single electron-on-helium device. (a) Optical image. (b) Circuit diagram. (c) Tilted, false-colored SEM image around the trap region. (d) Schematic cross-section of the trap region shown in (c) with filled superfluid He and a trapped electron. Adapted with permission from Ref.~\cite{koolstra2019coupling}. Copyright 2019, Springer Nature.}
	\label{Fig:GewinDevice}	
\end{figure*}

Theoretical calculation shows two primary sources of electron charge qubit's coherence loss to excitations in helium, when there are no external vibrations and superfluid He is in thermal equilibrium. One is through the decay into capillary waves on the He surface, known as ripplons, and the other is through the decay into phonons in the bulk.~\cite{dykman2003qubits,schuster2010proposal} The electron is about 11~nm above the He surface, which is much large than the amplitude of ripplon excitations, so its coupling to ripplons is tiny. The rate of direct emission into individual ripplon is suppressed by energy-momentum mismatch. So the decay into ripplons is dominated by second-order processes in which the electron interacts with two
nearly opposite-traveling ripplons simultaneously. But the estimated decay rate through this process is still less than 1~kHz. The coupling to bulk phonons is more prominent. An electron creates an electric field that polarizes helium, which acts back to the electron. Bulk phonons in helium modulate the helium density and thus the polarization, which changes the electron energy. The estimated decay rate through this mechanism is $\sim$30~kHz at $\sim$6~GHz qubit frequency. If this can be practically verified, then the strong coupling condition $g>\kappa,\gamma$ can be fulfilled and the electron's motional states can be controlled and readout by microwave photons. 

However, Monarkha pointed out in 1978 that the interaction of an electron with short-wavelength ripplons involved in the electron energy relaxation should be described differently from the conventional treatment which assumes an infinite surface barrier.~\cite{monarkha1978influence,monarkha2013two} In the perturbation theory, the main contribution to the interaction comes from the penetration of the electron wavefunction inside the liquid helium. With a finite surface barrier, Monarkha estimated the relaxation rate on the order 1~MHz for the excited Landau states in an external magnetic field and for the excited Rydberg states of quantized vertical motion.~\cite{monarkha1978influence,monarkha2007decay} Recently, this result has been experimentally confirmed for the relaxation of the excited Rydberg states.~\cite{kawakami2021relaxation} Although no explicit calculations have been presented for the relaxation of the excited states of electron lateral motion in a quantum dot, it is reasonable to expect that the corresponding rate is on the same order of magnitude.

Since around 2011, experimental effort has been made in coupling electrons on helium (eHe) with microwave photons in a cQED architecture. Fig.~\ref{Fig:GeDevice} (a-f) shows the first-generation devices from Schuster's group use a standard coplanar waveguide (CPW) resonator, where a single stripline is embedded between two ground planes and terminated at a half wavelength.~\cite{yang2016coupling} The microwaves are coupled in and out at the two ends where the electric field is maximal. The middle point corresponding to a quarter wavelength is a nodal point where the electric field is zero. This point is used to deliver a dc voltage by a T-structure, without interfering with the ac signal, to provide a trapping potential for the electrons in the channel. At the two ends of the resonator, there are additional dc electrodes running from the ground plane. They are designed to trap individual electrons and tune the transition frequencies around the resonator frequency. The resonator with a quality factor, $Q=10^5$, can precisely sense the filling process of superfluid He through the channel. The cell that hosts the sample chip has a millimeter-diameter pit to hold bulk superfluid He. When the pit is filled with He at certain level, the height difference between the channel on the chip and the pit determines at what level the channel can be filled, owing to the superfluid creeping effect and capillary action. He-filled channel changes the dielectric environment of the superconducting resonator. Therefore, with a controllable puff filling system, one can trace the channel filling status from empty to fully filled, by monitoring the resonator frequency shift. Since superfluid He wets almost any surfaces, one can perform a numerical simulation to find out the dielectric filling induced frequency shift on the resonator. As shown in Fig.~\ref{Fig:GeHeFilling} (a,b), the actual observation of resonator frequency shift turns out to be well consistent with the simulation. With the same device, repeated loading and unloading an ensemble of electrons in and out of the channel could be observed. The electrons are believed to form quasi-1D classical Wigner crystals in the channel. 

\begin{figure}[thb]
	\centering
	\includegraphics[scale=1.1]{\imgpath/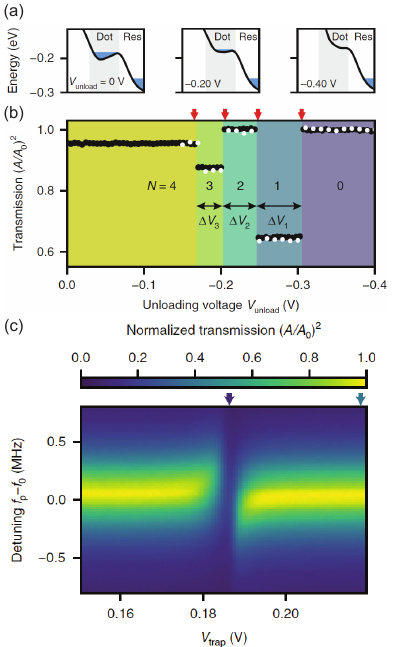}
	\caption{Resonator signatures of few-electron clusters. (a) Schematic of the unloading procedure. (b) Observed five distinct plateaus with decreasing voltages suggesting different number of electrons in the trap. (c) Single electron resonator spectroscopy showing a normalized transmission amplitude as function of trap voltage and microwave probe detuning. Adapted with permission from Ref.~\cite{koolstra2019coupling}. Copyright 2019, Springer Nature.}
	\label{Fig:GerwinCoupling}
\end{figure}

The second-generation devices from Schuster's group has a more sophisticated design, see Fig.~\ref{Fig:GewinDevice}, targeting trapping a single electron and coupling it with microwave photons.~\cite{koolstra2019coupling} As shown in Fig.~\ref{Fig:GewinDevice} (b,c), the resonator is made of a tuning-fork like quarter-wavelength double-stripeline resonator, embedded in an etched-down channel in Si, also clamped between ground planes. The dc voltages on the resonator is applied at the quarter-wavelength nodal point. An oval-shaped trap is made at the end of the tuning fork where the electric field is strongest. A separate trap line runs from the other side of the channel into the double-stripline resonator. It has a specifically designed cross shape in the trap region, which makes the electron more tightly confined in the direction along the channel and less confined in the direction across the channel. The cross-channel direction is aligned with the electric dipole orientation and the electric field direction of microwave photons. The differential mode with electric field pointing from one stripline to the other couples with the electric dipole transition of the electron (See Fig.~\ref{Fig:GewinDevice} (d).) Four additional dc electrodes are fabricated around the trap region for deterministically loading and unloading electrons and tuning their frequencies. These dc lines are accompanied with on-chip LC filters to reduce microwave leakage. (See Fig.~\ref{Fig:GewinDevice}~(a).) The experimental observation shows signatures of trapping several electrons in the trap region and one-by-one kicking them off the trap until only one electron is retained in the trap, see Fig.~\ref{Fig:GerwinCoupling}~(a,b). The key observation is the coupling (level splitting) between a single electron and microwave photons Fig.~\ref{Fig:GerwinCoupling}~(c). The coupling strength is $\sim$5~MHz. However, the electron linewidth is $\sim$80~MHz. This is much larger than the coupling strength, so does not satisfy the condition for single electron-photon strong coupling. A probable interpretation of the broadened electron linewidth than theoretical estimation is that the pulse tube operation of a closed-loop dilution refrigerator produces additional surface vibration on superfluid He and thus decoherence to the electron.

\subsection{Electron spin qubits on helium via circuit QED}

Schuster {\etal} also proposed an approach to realize eHe spin qubits in the cQED architecture.~\cite{schuster2010proposal} Natural helium contains only 1.37~ppm abundance of $^3$He. Hence superfluid $^4$He contains negligible nuclear spins around and is the cleanest natural spin bath for electron qubits. The spin coherence time is expected to be over 100~s.~\cite{lyon2006spin}

\begin{figure}[thb]
	\centering
	\includegraphics[scale=0.95]{\imgpath/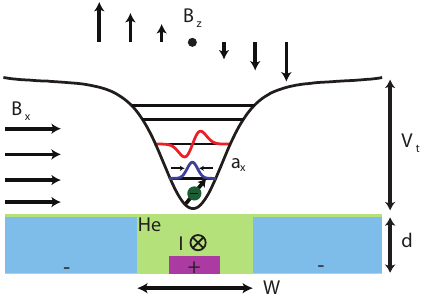}
	\caption{Side view of trap electrodes with energy levels and wavefunctions of electron motional states above the surface of liquid helium. A uniform magnetic field in the $x$ direction defines the spin quantization axis. A current is sent through the center electrode to creating a field gradient in $z$ to couple the motional and spin degrees of freedom. Adapted with permission from Ref.~\cite{schuster2010proposal}. Copyright 2010, American Physical Society.}
	\label{Fig:HeliumGradientField}
\end{figure}

However, direct coupling of a single electron's spin (magnetic dipole moment) with (magnetic part of) microwave photons in a resonator is only on the order of 10~kHz. One viable solution is to use the electric-dipole spin resonance (EDSR) to enhance the effective coupling strength.~\cite{kawakami2016gate,benito2017input,mi2018coherent} By introducing a synthetic spin-orbit (SO) coupling, the spin and motional states of the electron is hybridized, see Fig.~\ref{Fig:HeliumGradientField}. An in-plane uniform magnetic field $B_0$ of 0.2~T can be applied across the channel in the $x$ direction to define the spin-quantization axis. For Nb film, it has been shown that a resonator qualify factor $Q>20000$ can be maintained under this in-plane magnetic field, offering necessary sensitivity for spin qubits readout. An out-of-plane nonuniform magnetic field in the $z$ direction with a gradient along the $x$ axis, $\partial_xB_z$, can be generated by a current $I$ in the $y$ direction along the central stripline of the CPW resonator.~\cite{schuster2010proposal} This gives a synthetic SO-coupling term, $\hat{H}_{\SSS\text{SO}} = -2\muB (\partial_xB_z) \hat{x} \hat{s}_z$, in the qubit Hamiltonian, where $\muB$ is the Bohr magneton. When the electron is coupled with the electric field of microwave photons through the motional (charge) states, the enhanced effective coupling strength between the spin and photon can be approximated as
\begin{equation}
	g_s = \muB a_x \left(\partial_x B_z\right) \frac{g\sqrt{2}}{\hbar\omega_x (1-\omega_{\SSS\text{L}}^2/\omega_x^2)},
\end{equation}
where $\omega_{\SSS\text{L}} = 2\muB/\hbar$ is the Lamor frequency of the electron, $\omega_x$ is the charge qubit frequency in the harmonic trap approximation, $a_x = \sqrt{\hbar/m_e\omega_x}$ is the charge trap width, and $g$ is the original charge-photon coupling strength. The above expression holds when the $\omega_{\SSS\text{L}}$ is sufficiently detuned from the $\omega_x$. Assuming the current $I\sim 1$~mA at a channel depth $d=500$~nm away from the electron, the field gradient can be $\partial_xB_z\sim 8$~mG/nm. If $g=20$~MHz, $\omega_{\SSS\text{L}}-\omega_x=30$~MHz, then the effective $g_s\approx 0.5$~MHz. This will make $g_s>\kappa, \gamma$ in the strong coupling regime.

If the current is kept on and the SO coupling is kept on, then the overall coherence is a hybrid between the charge and spin coherence. The electron spin coherence alone can be over 1~s on liquid helium. When the SO coupling is on, the charge decoherence affects the overall decoherence and may bring it to 10~ms order. However, in principle, there is no need to keep the SO coupling on. To gate a spin qubit, a strong enough microwave pulse can provide the necessary gate between 0 and 1 spin states. The SO coupling is only needed during the readout. Therefore, it is more meaningful to use this controllable coupling only for spin-to-charge conversion before a charge readout. A theoretical calculation shows that a spin-to-charge conversion only needs a few nanosecond. After the charge readout, the spin states can be inferred from the charges states. This approach can push the qubit coherence time toward the theoretical spin coherence limit. 

At present, the experimental realization of eHe spin qubits through the cQED architecture is an active research topic. However, since (some kind of) spin-to-charge conversion is still practically necessary, vibrations of liquid He surface could still rapidly decohere charge states and impose a big challenge for the realization (at least the readout) of spin qubits.

\begin{figure*}[hbt]
	\centering
	\includegraphics[scale=1]{\imgpath/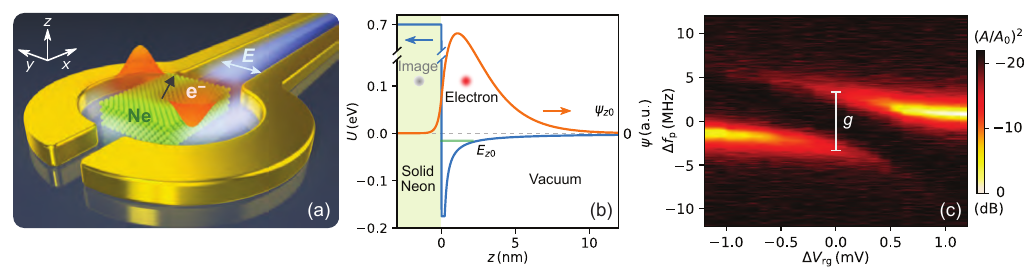}
	\caption{Schematic and properties of the electron-on-solid-neon (eNe) system.~\cite{zhou2022single} (a) Illustration of the eNe qubit platform based on the charge and spin states of a single electron trapped on the surface of solid Ne and manipulated by microwave photons in a superconducting quantum circuit. (b) Potential energy seen by an excess electron approaching a flat solid Ne surface and calculated ground-state eigenenergy and wavefunction in the out-of-plane ($z$) direction. (c) Strong coupling (vacuum Rabi splitting) between a single electron qubit and microwave photons in an on-chip superconducting resonator. Adapted with permission from Ref.~\cite{zhou2022single}. Copyright 2022, Springer Nature.}
	\label{Fig:AllElectronNeon}
\end{figure*}

\begin{figure*}[htb]
	\centering
	\includegraphics[scale=0.95]{\imgpath/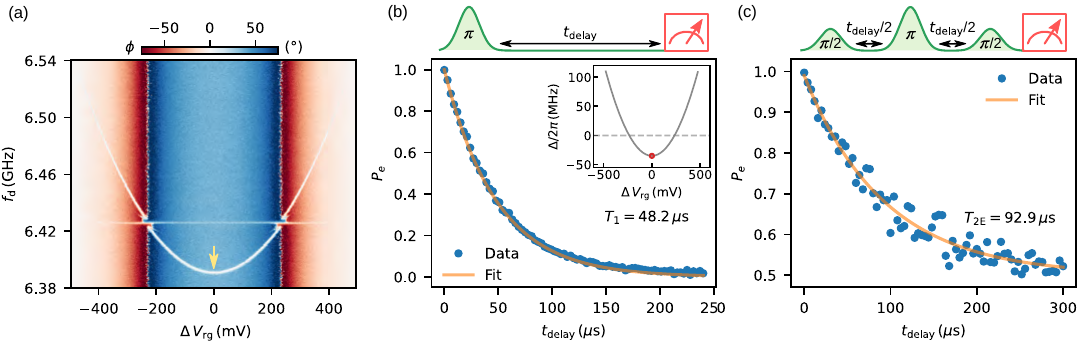}
	\caption{Experimental results on electron-on-solid-neon (eNe) qubits.~\cite{zhou2022single,zhou2024electron} (a) Qubit spectrum showing a quadratic shape with a charge-insensitive sweet spot. (b) Relaxation time measurement of the qubit on the sweet spot. (c) Hahn-echo coherence time measurement on the sweet spot. Adapted with permission from Ref.~\cite{zhou2024electron}. Copyright 2024, Springer Nature.}
	\label{Fig:SpectrumAndTimes}
\end{figure*}

\section{Quantum Electronics on Solid Neon}

\subsection{Electron charge qubits on neon via circuit QED}

\begin{figure*}[htb]
	\centering
	\includegraphics[scale=0.95]{\imgpath/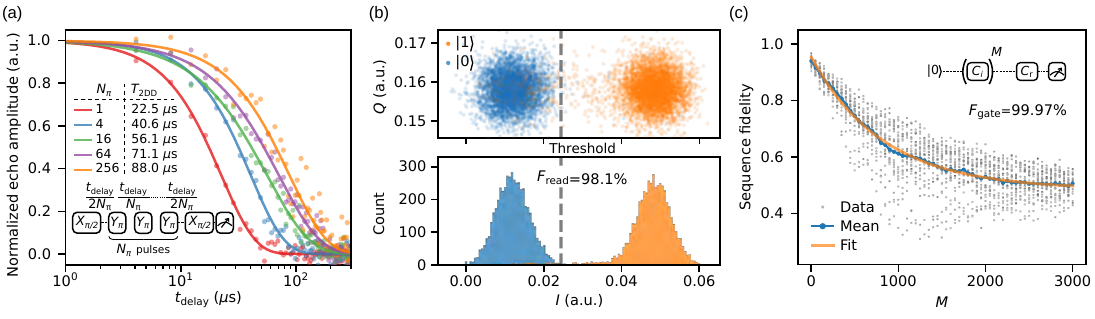}
	\caption{Experimental results on electron-on-solid-neon (eNe) qubits.~\cite{zhou2022single,zhou2024electron}. (a) Normalized echo amplitude versus total delay time for different number of dynamical decoupling pulses. (b) Single-shot readout fidelity without relying on a quantum-limited amplifier. (c) One-qubit gate fidelity using Clifford-based randomized benchmarking. Adapted with permission from Ref.~\cite{zhou2024electron}. Copyright 2024, Springer Nature.}
	\label{Fig:Fidelities}
\end{figure*}

In 2021, Zhou {\etal} achieved the first electron qubit in the QLS system by trapping and manipulating a single electron on a solid Ne (instead of a superfluid He) surface, see Fig.~\ref{Fig:AllElectronNeon} (a).~\cite{zhou2022single} Neon is the second noble element after He in the periodic table. It spontaneously solidifies below $\sim$24~K (triple-point temperature) and fundamentally removes the disadvantage of surface vibration of liquid He. Compared with conventional solid-state substrates (like Si and sapphire), solid Ne is much cleaner without chemical dangling bonds or two-level-system (TLS) fluctuators in the experimentally relevant frequency range. 

While the eNe and eHe systems look similiar, there are crucial differences. The Pauli barrier is lower and the polarization attraction is stronger for solid Ne. Therefore, the trapped electron is only 1-2~nm from a solid Ne surface, see Fig.~\ref{Fig:AllElectronNeon}(b) and Table~\ref{Table2}, based on the numerical solutions of the Schr\"odinger equation (the analytical solution assuming an infinite barrier overestimates the distance).~\cite{cole1969image,zhou2022single} This short distance makes the electron wavefunction more tightly attach to, and more strongly interact with the topography of the solid Ne surface. Besides, while it is known that liquid Ne wets almost all materials at its triple point, the actual growth of solid Ne thin film on a cQED chip during the continued cooldown is much harder to predict than superfluid He.

The cell and device are the same as those used in the eHe experiments, see Fig.~\ref{Fig:GewinDevice} (a-f),~\cite{koolstra2019coupling} with the main difference of replacing the liquid He layer with solid Ne. The microwave resonator is still the tuning-fork like quarter-wavelength double-stripline resonator. The trapping potential can be tuned by multiple dc electrodes, each of which has an on-chip low-pass LC filter to avoid microwave leakage outward. Ultralow-noise dc voltages are delivered on to the dc electrodes by first passing through thermocoaxes with 100~MHz cutoff, then pi-filters with 10~MHz cutoff, and then homemade RC filters with 10~Hz cutoff. Ne was filled at its liquid phase, then was solidified by cooling the device to below 24~K. There is no clear clue on how uniform the Ne coating is. Assuming the solid Ne conformally coats the resonator, based on the observed resonant frequency shift and the finite-element simulation, the thickness would be only about 10~nm. However, it is possible that more Ne is frozen in the trap region and less in the long resonator region. Controlled growth of solid Ne film on either flat substrates or patterned chips is under active development by Jin's group at this time. 

In their first series of experiments, strong coupling (vacuum Rabi splitting) between the charge (motional) states of an electron and microwave photons in an on-chip superconducting resonator at $\sim$6~GHz frequencies was achieved, see Fig.~\ref{Fig:AllElectronNeon} (c).~\cite{zhou2022single} The measured coupling strength $g$ is about 3.5~MHz, already greater than the electron linewidth about 1.7~MHz. A 2-tone qubit spectroscopy measurement was performed and shows a quadratic charge-qubit spectrum that is very similar to a semiconductor double-quantum-dot (DQD) qubit, see Fig.~\ref{Fig:SpectrumAndTimes} (a). Rabi oscillations and dispersive readout are also demonstrated. Their first set of measurements without particularly driving the electron at the charge-insensitive sweet spot gives a Ramsey coherence time $T_2^*$ of 50~ns and a Hahn echo coherence time $T_{2\text{E}}$ of 220~ns. These results are already better than all the traditional semiconductor and superconducting charge qubits.

In their second series of experiments, they managed to refine their Ne growth procedure and trap an extremely stable and long-lived electron.~\cite{zhou2024electron} While the electron-photon coupling strength is only 2.3~MHz, the electron linewidth at the charge sweet spot drops below 0.1~MHz. The observed relaxation time $T_1$ and coherence time $T_2$ both have reached the order of 0.1~ms. The $T_1$ is only limited by the Purcell enhancement to the spontaneous emission of photons into the cavity. On the charge sweet spot, $T_1=43$~{\textmu}s and $T_{2\text{E}}=93$~{\textmu}s, which is approximately $2T_1$, meaning the high-frequency decoherence is almost solely caused by radiative relaxation, see Fig.~\ref{Fig:SpectrumAndTimes} (b,c). Away from the charge sweet spot, $T_1$ can go above $100$~{\textmu}s. Theoretically, if the Purcell effect can be suppressed by choosing a large qubit-resonator detuning, then the leading relaxation and decoherence mechanism comes from bulk phonons. Our estimated phonon-limited coherence time can be on the order of 1--10~ms, suggesting plenty of room to improve for our current qubits. Moreover, on another qubit that is slightly less coherent, when $T_{2\text{E}}$ does not yet reach $2T_1$,  the coherence time at the charge sweet spot can be significantly extended by implementing the dynamical-decoupling (DD) pulse sequences. It suggests that the major sources of decoherence for the specific electron qubit are low-frequency noises.

\begin{figure}[htb]
	\centering
	\includegraphics[scale=0.95]{\imgpath/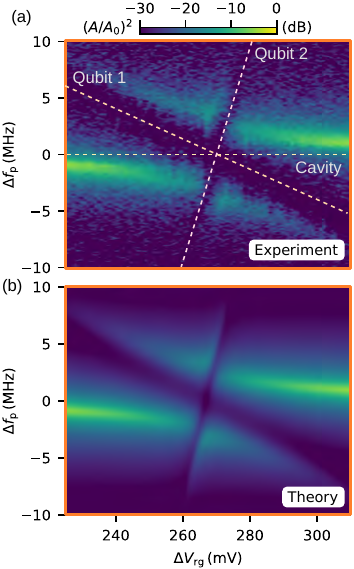}
	\caption{Spectroscopic characterization of two eNe charge qubits coupled to a common resonator. a. Experimental observation. b. Theoretical calculation. Adapted with permission from Ref.~\cite{zhou2024electron}. Copyright 2024, Springer Nature.}
	\label{Fig:TwoQubit}
\end{figure}

The readout and gate fidelities of the eNe qubit were also characterized, see Fig.~\ref{Fig:Fidelities} (a-c). The single-shot readout fidelity without relying on a quantum-limited amplifier (QLA) is measured to be 98.1\%. If a QLA, such as a traveling-wave parametric amplifier (TWPA) or a Josephson parametric amplifier (JPA), is used, the readout fidelity should go above 99\% with a shorter readout pulse $\sim$100~ns. The one-qubit gate fidelity calibrated by the Clifford-based randomized benchmarking technique is also measured to be 99.97\%, which is comparable to the state-of-the-art superconducting transmon qubits. Simultaneous strong coupling of two qubits with a common bus resonator has also been demonstrated, as a first step toward two-qubit entangling gates for universal quantum computing.~\cite{zhou2024electron} The experimental observation and theoretical calculation of the coupling of two qubits to the same resonator show excellent agreement, see Fig.~\ref{Fig:TwoQubit} (a,b). All these results manifest that the eNe charge qubits have outperformed all the traditional charge qubits and rivaled the best superconducting transmon qubits to date.~\cite{blais2021circuit,stano2022review} This endeavor has accomplished the two-decade dream of using QLS to host long-coherence high-fidelity electron qubits~\cite{platzman1999quantum,dykman2003qubits,lyon2006spin,schuster2010proposal,yang2016coupling,koolstra2019coupling,jin2020quantum,chen2022electron,dykman2023spin,kawakami2023blueprint}. 

Most recently, two-qubit coupled devices with improved design have been fabricated and measured, targeting the realization of two-qubit gates. Since the electrons in this system have comparatively small electric dipole moments, the most critical step is to enhance the electron-photon coupling strength by confining photon more strongly. High-KI TiN films have been used to replace Nb and have enhanced the coupling strength to $\sim$10~MHz range. So long as the electron linewidth of two qubits can be maintained at the 0.1~MHz level,  two-qubit gates should be achievable in the near term.

\subsection{Electron spin qubits on neon via circuit QED}

Chen {\etal} calculated the spin coherence time of a single electron on a solid Ne surface.~\cite{chen2022electron} Natural Ne consists of three stable isotopes: $^{20}$Ne (90.48\%), $^{21}$Ne (0.27\%), and $^{22}$Ne (9.25\%) with the abundance of each component given in the parentheses. $^{20}$Ne and $^{22}$Ne have 0 nuclear spin while $^{21}$Ne has $\frac{3}{2}$ nuclear spin~\cite{hubbs1956spin}. All Ne atoms in the ground state have closed shells and fully paired electrons. The total angular momentum of the shell electrons is zero and hence does not produce intrinsic magnetic moment. 

The magnetic response of a Ne atom is to the leading order diamagnetic and is a quantum mechanical effect. The induced magnetization energy is proportional to the square of applied magnetic field and always increases with the field strength irrespective of the field direction~\cite{ashcroft2022solid}. In our case, both the electron and solid Ne experience a constant external magnetic field $B_{0}\sim0.2$~T. To be compatible with the superconducting devices, this field should be applied along the $x$ direction that is parallel to the superconducting films and solid Ne surface. It magnetizes the Ne sample through the diamagnetism of Ne atoms. The induced magnetization generates a magnetization surface current. The magnetization current then generates a magnetic field that acts on the spin of electron. During this process, the thermal fluctuations of bulk phonon modes in the solid Ne change the Ne mass density and consequently change the volume magnetic susceptibility. This temporal variation leads to a fluctuating magnetization current and thus a fluctuating magnetic field that acts on the electron. This mechanism induces spin relaxation and decoherence~\cite{dykman2023spin}. The calculated relaxation and coherence times through this mechanism are longer than $10^6$~s and so are not the limiting factor.

The electron-nuclear spin-spin interaction is the more dominant decoherence mechanism. Ne has 2700~ppm of $^{21}$Ne. Under a $B_0$ field of $\sim$0.2~T, Ne nuclear spin resonance frequency is 4.4~MHz, which allows much thermal population even at 10~mK temperature. Taking the secular approximation for the hyperfine interaction and Gaussian distribution of the random Overhauser field, the inhomogeneous dephasing time of the electron spin is $T_2^*=0.16$~ms. However, dynamical decoupling can significantly extend the coherence by removing the MHz low-frequency noise from the nuclei. This can lead to the Hahn echo coherence time $T_2 = 30$~ms. Practically, the influence of $^{21}$Ne nuclear spins on the electron spin coherence can be suppressed by isotopic purification. Isotopically purified $^{22}$Ne with only 1~ppm of $^{21}$Ne is commercially available (Cryoin Engineering Ltd.)~\cite{bondarenko2018multi} For 1~ppm of $^{21}$Ne, the estimated inhomogeneous dephasing time $T^{*}_{2}$ is $0.43$~s, and the coherence time $T_2$ under Hahn echoes can reach $81$~s.~\cite{chen2022electron}

Experimental realization of eNe spin qubits can follow the same EDSR scheme as that of the envisioned eHe spin qubits. Nonetheless, solid Ne may provide an additional advantage of being able to host more tightly confined electrons than on liquid He to potentially achieve DQD based spin qubits or even spin singlet-triplet (ST$_0$) qubits.~\cite{burkard2023semiconductor}

\subsection{Quantum ring states of electrons on solid neon}\label{Sec-eNe-ring}
Despite the demonstrated exceptional performance of the eNe qubits, recent experiments have also unveiled some intriguing phenomena. For instance, it was observed that when the electric trapping potential was reduced, the shift in the excitation spectrum associated with the electron's lateral motion was significantly less than expected.~\cite{zhou2024electron} Moreover, in some experimental runs, the electrons remained anchored to the Ne surface even after removing the trapping potential entirely. These observations suggest the existence of an alternative mechanism confining the electron laterally on the neon surface. Indeed, earlier studies on the mobility of electrons trapped on solid H$_2$ also revealed that electrons could become immobile on rough H$_2$ surfaces.~\cite{kono1991surface1,kono1991surface2,albrecht1990surface} In a recent theoretical work, Kanai {\etal} explored the interaction between an electron and an isolated surface topography, such as a bump or a valley on a solid Ne.~\cite{kanai2024single} These surface features can spontaneously form due to the Stranski-Krastanov growth mode of solid Ne at temperatures below its triple point.~\cite{esztermann2002triple,shchukin2013formation} It was revealed that the electron can form localized quantum ring states around the surface bump with properties aligning well with the experimental observations.

\begin{figure}[h]
	\includegraphics[width=\columnwidth,keepaspectratio]{\imgpath/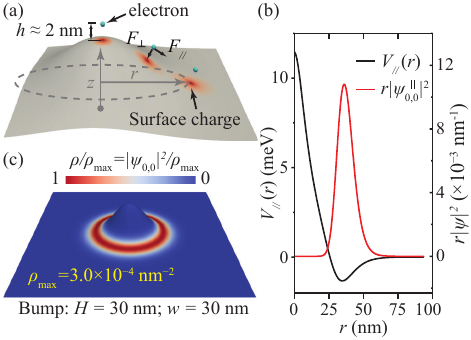}
	\caption{\label{Wei-Fig3} (a) A schematic showing the image charge induced by an electron bound to a solid Ne surface bump. (b) Lateral potential energy $V_\parallel(r)$ of the electron on a representative solid Ne surface bump with $H=30$~nm and $w=30$~nm. (c) The profile of the electron's ground-state eigenfunction. Adapted with permission from Ref.~\cite{kanai2024single}. Copyright 2024, American Physical Society.}
\end{figure}

Consider an electron bound to a flat solid Ne surface, where the resulting image charge symmetrically distributes around the electron's vertical projection point on the surface. The electron experiences a perpendicular force $F_\perp$, pulling it towards to the Ne surface, with no net force $F_\parallel$ parallel to the surface. On the other hand, when the electron is placed on a curved surface, such as a Ne surface bump as depicted in Fig.~\ref{Wei-Fig3} (a), the induced image charge can exhibit a nonsymmetric distribution around the electron's projection point. This asymmetry results in a residue $F_\parallel$ along the surface. For surface bumps or valleys with heights $H$ and half-widths $w$ significantly larger than the distance between the electron and the Ne surface (i.e., $\langle z\rangle\simeq2$~nm), the changes in image charge distribution and the resultant $F_\perp$ are minimal as compared to those on a flat Ne surface.~\cite{kanai2024single} Therefore, the electron remain bound at $\langle z\rangle\simeq2$~nm above the Ne surface regardless the underlying surface profile. To study the electron's lateral motion, one may integrate $F_\parallel$ along the curved surface to derive a lateral potential energy $V_\parallel(r)$. Fig.~\ref{Wei-Fig3} (b) shows the obtained $V_\parallel(r)$ for a representative bump with $H=30$~nm and $w=30$~nm. Notably, $F_\parallel$ changes sign around the waist of the bump, leading to a quantum-ring trapping potential encircling the bump with a potential depth of about $-1.33$~meV. This potential depth is large enough to trap the electron without any externally applied potential. The electron's eigenstates can be determined by solving the Schr\"{o}dinger equation on the curved Ne surface with the derived potential $V_\parallel(r)$. Fig.~\ref{Wei-Fig3} (c) shows the profile of the electron's ground-state wavefunction, which aligns with the trapping potential $V_\parallel(r)$. This study also reveals that surface valleys repel electrons at large distances due to their reversed lateral potential profiles as compared to surface bumps.~\cite{kanai2024single} For an electron bound in the quantum ring ground state of zero angular momentum, an oscillating in-plane electric field produced by the resonator photons can bring it to an excited state of nonzero angular momentum. The calculation shows that the transition frequency is primarily controlled by $w$, or equivalently, the circumference $\pi w$. For a bump with $w\simeq30$~nm, the transition frequency matches well the resonator's photon frequency.~\cite{kanai2024single}

\section{Outlook}

\subsection{Electron qubits on solid hydrogen}

In addition to solid Ne, solid H$_2$ is another candidate to support solid-state electron qubits. It is known that the hovering distance of an electron on solid H$_2$ (eH$_2$) is nearly the same as that of an eNe. (See Table~\ref{Table2}.) However, solid H$_2$ has a higher Pauli barrier and so the penetration depth of the electron wavefunction into solid H$_2$ is less than that into solid Ne. This may suggest reduced influence of surface roughness on electron trapping and transfer. Moreover, due to the lower molecular mass, solid H$_2$ has a much higher zero-point motion than solid Ne, as manifested by a larger de Boer parameter. (See Table~\ref{Table1}.) The triple-point temperature of H$_2$ is much lower than Ne. These may assist the natural formation of smoother surfaces on solid H$_2$ than on solid Ne.

Unlike liquid He or solid Ne, solid H$_2$ and D$_2$ are molecular crystals. They are constructed from indistinguishable nuclei and possess an ortho-para molecular wavefunction symmetry.~\cite{silvera1980solid} The ortho-para transition is forbidden for isolated molecules and takes place in solid H$_2$ and D$_2$ with a rate of 1.9\%/h and 0.06\%/h, respectively. This results in a very poor thermalization of H$_2$ and D$_2$ solids upon cooldown. The ongoing ortho-para conversion is accompanied by a significant heat release (170.5~K/molecule for H$_2$ and 86~K/molecule for D$_2$) which can significantly disturb qubit operations. This may be avoided by using para-H$_2$ instead of the normal (75\% ortho- and 25\% para-)H$_2$. For nm-thick films, ortho-para conversion can be accelerated by paramagnetic species or radicals always present on surfaces. Nonetheless, a much faster thermalization is expected in solid HD, in which unlike solid H$_2$ and D$_2$, rotational transitions are not hindered by the molecular wavefunction symmetry.

\subsection{Electronic structures at the interface of solid neon and superfluid helium}

Previous studies of quantum electronics on QLSs mainly focused on homogeneous QLS species, \eg, electrons on purely superfluid He, solid H$_2$, or solid Ne exposed to a vacuum. No systematic studies have been conducted to the quantum electronics on the top surface or at the interface of a multilayer heterogeneous mixture, such as that of liquid He, solid H$_2$, and solid Ne. Heterogeneous QLSs can host extraordinary electronic structures and enable large device functionalities that are of both fundamental interest and practical applications. We envision these topics to constitute some of the future directions in this area. 

Jin theoretically predicted that at the interface of solid Ne and superfluid He, a single electron forms a self-confined dome structure, in which the flat side attaches to the solid Ne and curved side dips into the superfluid He by several nanometers.~\cite{jin2020quantum} This electron-dome structure may be viewed as a deformed electron-bubble structure in bulk liquid or solid $^3$He and $^4$He.~\cite{sommer1964liquid,fowler1968electronic,kajita1983stability,grimes1990infrared,golov1995spectroscopic,guo2009experiments} Jin also showed that many such electron domes can form a classical Wigner crystal that resembles a quantum-dot array. This array
can exhibit the quantum optical phenomenon of superradiance in the mid-infrared wavelength regime.

\begin{figure}[htb]
	\centerline{\includegraphics[scale=0.8]{\imgpath/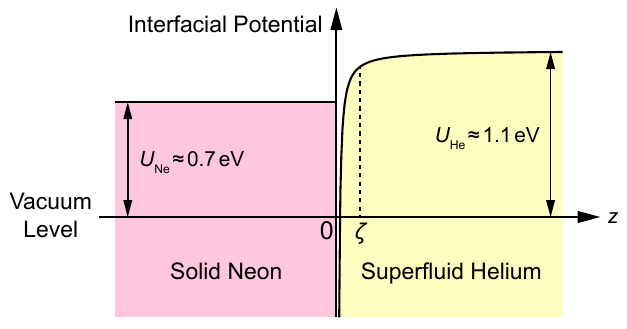}}
	\caption{Interfacial potential for an electron sandwiched between flat superfluid He and solid Ne at zero pressure. Adapted with permission from Ref.~\cite{jin2020quantum}. Copyright 2020, IOP Publishing.}\label{Fig:InterfacialPotential}
\end{figure}

A bosonic density functional theory (DFT) is used to calculate the ground state, excited states, and optical transitions of these extraordinary states.~\cite{otten2020coherent} Fig.~\ref{Fig:InterfacialPotential} gives the DFT calculated interfacial potential seen by an electron sandwiched between flat solid Ne and superfluid He. It consists of three contributions: the Pauli-exclusion potential barrier from Ne and He, respectively, and the image-charge attractive potential from Ne. It can be approximated as
\begin{equation}
	V(z) =
	\begin{cases}
		V_{\text{Ne}}, &\  z<0, \\
		\DS -\frac{\epsilon_{\text{Ne}}-1}{\epsilon_{\text{Ne}}+1}\frac{e^2}{4z} + V_{\text{He}} \tanh^2\left(\frac{z}{\zeta}\right), &\  z>0. \label{Eqn:Potential}
	\end{cases}
\end{equation}
Here, $V_{\text{He}}\approx 1.1$~eV and $V_{\text{Ne}}\approx0.7$~eV are the bulk Pauli barriers of superfluid He and solid Ne to the electron, and $\epsilon_{\text{Ne}} = 1.244$ is the dielectric constant of solid Ne.~\cite{cole1969image} Since the polarizability of He is much smaller compared with Ne, we can completely ignore its effect~\cite{maris2003properties,jin2010vortex} and simply take its dielectric constant as that of vacuum, $\epsilon_{\text{He}} = 1.056\approx 1$. $\zeta\approx 1$~{\AA} is the characteristic length within which the helium density sharply increases from zero to its bulk value.~\cite{jin2010vortex} Such a continuous approximation may not be accurate microscopically.~\cite{krotscheck1985liquid}
Among all the parameters above, only $V_{\text{He}}$ varies appreciably with pressure $p$ (or equivalently, helium number density $n$), and this pressure dependence is naturally included in the DFT calculation. Unlike a Ne-vacuum interface,~\cite{kajita1984new} the strong repulsion from He overrides the weak and long tail of polarization potential from Ne. Therefore, an attractive potential only exists in the $\sim$1~{\AA} thick region $0<z<\zeta$. The electronic structure is dominated by the repulsive barriers from the bulk Ne and He and is insensitive to the exact profile of attractive polarization potential or interfacial structure within the $0<z<\zeta$ region, as exemplified in Eq.~(\ref{Eqn:Potential}) and Fig.~\ref{Fig:InterfacialPotential}. 

\begin{figure}[hbt]
	\centerline{\includegraphics[scale=0.8]{\imgpath/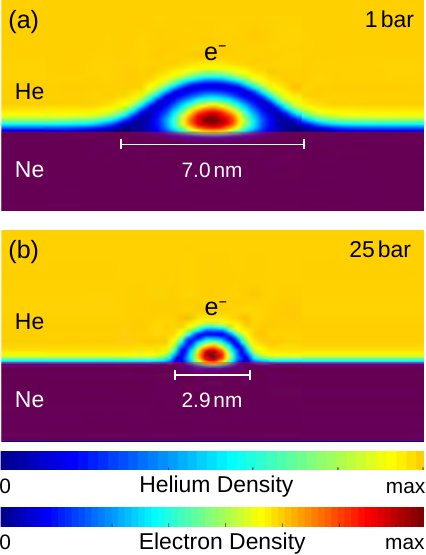}}
	\caption{Calculated ground-state electron and helium density profiles of the single-electron dome structure at the interface between superfluid He and solid Ne under 1 to 25~bar pressure (a-b). The outside of the dome shows the helium density and the inside shows the electron density; the area of overlap is negligible in the plot. Adapted with permission from Ref.~\cite{jin2020quantum}. Copyright 2020, IOP Publishing.}\label{Fig:BubbleImages}
\end{figure}

In Fig.~\ref{Fig:BubbleImages} (a,b), the DFT calculations shows that increasing pressure from 1 to 25~bar (below the liquid-solid phase transition of $^4$He) squeezes the dome diameter $D$ from 7~nm to 2.9~nm and dome height $H$ from 2.15~nm to 1.43~nm. This is in contrast to the electronic structure on the traditional He/Ne-to-vacuum surface, where pressure can only be zero due to the vacuum and cannot serve as a tuning knob in confining the electron. Correspondingly, the optical transition wavelengths are in the mid-infrared (mid-IR) regime from 7.66~\textmu m to 24.3~\textmu m wavelength.~\cite{jin2020quantum} 

When a number of electrons are deposited at the interface and bounded by a hard-wall potential, they can form a classical Wigner crystal. The electron density can be higher than $3\times 10^{10}$~cm$^{-2}$.
Such a Wigner crystal is equivalent to a highly compact quantum-dot (QD) array. 
The distance between the QDs can be $<$100~nm, which is much shorter than the mid-IR wavelength of $\sim$10~\textmu m. The fluorescence behavior of this QD array has reached the condition of superradiance.~\cite{dicke1954coherence,ernst1968emission,rehler1971superradiance,scheibner2007superradiance}
All the electrons coherently interact with the same photon field, and exhibit intensity-enhanced and
lifetime-shortened emission, drastically different from the spontaneous emission of a single or a sparse
ensemble of electrons. While superradiance in the visible regime has been experimentally observed, it has not been observed in the mid-IR regime because of lack of appropriate emitters. It would be appealing to realize superradiance in a purely electronic crystal. 

\subsection{Electron qubits on the surface of heterogeneous quantum liquids and solids}

Heterogeneous quantum liquids and solids may improve the performance of electron qubits that are trapped or transferred on the top surface.

For instance, one can first coat the conventional substrate, \eg, Si or sapphire, with a thin layer ($\sim$100~nm) of solid Ne and then cover the solid Ne with a thin layer ($\sim$10~nm) of superfluid He. An electron qubit hovers above the superfluid He in the vacuum. On the one hand, the thicker solid Ne serves as a vibration stabilizer for the superfluid He film and, meanwhile, a decoherence mitigator to prevent the TLS fluctuators or quasiparticles in the substrate from harming the qubit coherence. On the other hand, the superfluid He serves as a surface smoother to cue the potentially rough solid Ne surface and enhance the mobility when transferring an electron spin qubit above. Nonetheless, prior studies of electron mobility on an ultrathin (a few atomic layers of) He film covering solid hydrogen (whose properties are similar to solid neon) showed a decrease, instead of an increase, of the electron mobility.~\cite{paalanen1985electron,cieslikowski1987investigation} This is interpreted as additional scattering from the density fluctuations of the unsaturated topmost helium atomic layer. With further increased He film thickness, additional scattering from superfluid ripplons can also set in. It is unclear yet whether there exists an optimal thickness for superfluid He on solid H$_2$ or Ne, so that the electron mobility can be improved and the surface vibration remains suppressed.

\begin{figure}[hbt]
	\includegraphics[scale=0.7]{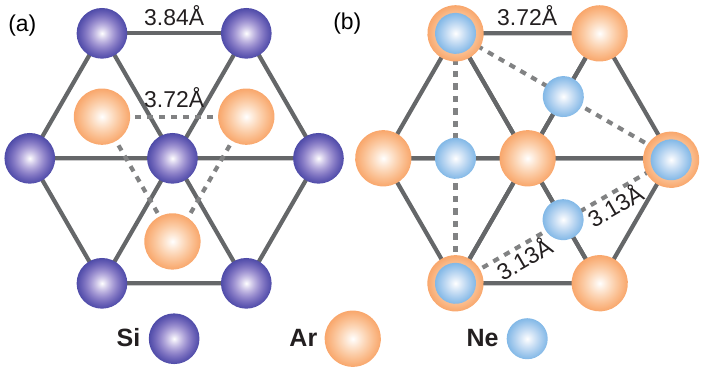}
\caption{Conceived nearly perfect lattice matching. (a) Ar-$\langle111\rangle$ on Si-$\langle111\rangle$. (b) Ne-$\langle111\rangle$ on Ar-$\langle111\rangle$.}
\label{Fig:LatticeMatching}
\end{figure}

Moreover, it may be useful to introduce a classical noble-element solid, such as solid argon (Ar), as a lattice-matching layer in this system, see Fig.~\ref{Fig:LatticeMatching} (a,b). Si has a diamond-cubic crystal structure with a square lattice constant $a_{\SSS\text{Si}\square}=5.43$~{\AA} on its $\langle100\rangle$ plane and a triangular lattice constant $a_{\SSS\text{Si}\triangle}=3.84$~{\AA} on its $\langle111\rangle$ plane. This triangular lattice is commensurate with a $30^\circ$-rotated triangular lattice with a lattice constant $\sqrt{3}a_{\SSS\text{Si}\triangle}=6.65$~{\AA} and a twice-larger triangular lattice constant $2a_{\SSS\text{Si}\triangle}=7.68$~{\AA} on the same $\langle111\rangle$ plane. Under zero pressure, both solid Ne and solid Ar have a face-center-cubic (fcc) crystal structure. On their $\langle100\rangle$ plane, their square lattice constant is $a_{\SSS\text{Ne}\square}=4.43$~{\AA} and $a_{\SSS\text{Ar}\square}=5.26$~{\AA}, respectively. On their $\langle111\rangle$ plane, their triangular lattice constant is $a_{\SSS\text{Ne}\triangle}=3.13$~{\AA} and $a_{\SSS\text{Ar}\triangle}=3.72$~{\AA}, respectively.~\cite{Klein1976RareGas,ebner1979density} Hence a twice-large Ne-$\langle111\rangle$ and a $30^\circ$-rotated Si-$\langle111\rangle$ has a (triangular) lattice misfit of $1-2a_{\SSS\text{Ne}\triangle}/\sqrt{3}a_{\SSS\text{Si}\triangle} = 5.9\%$,  a Ar-$\langle111\rangle$ and a Si-$\langle111\rangle$ has a (triangular) lattice misfit of $1-a_{\SSS\text{Ar}\triangle}/a_{\SSS\text{Si}\triangle} = 3.1\%$, a twice-large Ne-$\langle111\rangle$ and a $30^\circ$-rotated Ar-$\langle111\rangle$ has a mutual (triangular) misfit of only $1-(2a_{\SSS\text{Ne}\triangle}/\sqrt{3}a_{\SSS\text{Ar}\triangle}) = 2.8\%$. These lattice misfits are quite small, considering that the soft van der Waals interaction between noble-element atoms allows much easier stress relief than conventional solids.~\cite{sohaili2005triple} Therefore, it is conceivable to obtain, first of all, lattice-matched solid Ar-$\langle111\rangle$ on Si-$\langle111\rangle$ within the 3.1\% misfit, and then lattice-matched solid Ne-$\langle111\rangle$ on Ar-$\langle111\rangle$ within the 2.8\% misfit.

In summary, all the ideas and concepts envisioned above are worth systematic future exploration.

\section*{Acknowledgements}

D. J. and W. G. acknowledge support from the National Science Foundation (NSF) under Award No. OSI-2426768. D. J. acknowledges support from the Air Force Office of Scientific Research (AFOSR) under Award No. FA9550-23-1-0636 and the Julian Schwinger Foundation for Physics Research under Award No. JSF-20-07-0001. W. G. acknowledges support from the National Science Foundation (NSF) under Award No. DMR-2100790 and the Gordon and Betty Moore Foundation through Grant GBMF11567. W. G.'s work was conducted at the National High Magnetic Field Laboratory at Florida State University, which is supported by the National Science Foundation Cooperative Agreement No. DMR-2128556 and the state of Florida.
D. K. acknowledges support from the Okinawa Institute of Science and Technology (OIST) Graduate University and the Grant-in-Aid for Scientific Research (Grant No. 23H01795 and 23K26488) KAKENHI MEXT.

\bibliography{AllRef}

\end{document}